\documentclass{JINST}

\usepackage[utf8]{inputenc}
\usepackage{graphicx}
\usepackage{amsmath}
\usepackage[english]{babel}
\usepackage{tabularx}		
\usepackage{cite}
\usepackage{url}

\newcommand{\geant}{{\sc Geant4}}

\graphicspath{{figures/}}

\title{The Time Structure of Hadronic Showers in highly granular Calorimeters with Tungsten and Steel Absorbers}

\author{\centering 
\LARGE\bf The CALICE Collaboration
}

\author{\centering
C.\,Adloff, 
J.-J.\,Blaising, 
M.\,Chefdeville, 
C.\,Drancourt,
R.\,Gaglione, 
N.\,Geffroy, 
Y.\,Karyotakis, 
I.\,Koletsou, 
J.\,Prast,
G.\,Vouters 
\\ \it
Laboratoire d'Annecy-le-Vieux de Physique des Particules, Universit\'{e} de Savoie,
CNRS/IN2P3,
9 Chemin de Bellevue BP110, F-74941 Annecy-le-Vieux CEDEX, France
}

\author{\centering
J.\,Repond, 
J.\,Schlereth, 
L.\,Xia 
\\ \it
Argonne National Laboratory,
9700 S.\ Cass Avenue,
Argonne, IL 60439-4815,
USA}

\author{\centering
E.\,Baldolemar, 
J.\,Li$^a$, 
S.\,T.\,Park, 
M.\,Sosebee, 
A.\,P.\,White, 
J.\,Yu 
\\ \it
Department of Physics, SH108, University of Texas, Arlington, TX 76019, USA
}

\author{\centering
G.\,Eigen 
\\ \it
University of Bergen, Inst.\, of Physics, Allegaten 55, N-5007 Bergen, Norway
}

\author{\centering 
M.\,A.\,Thomson, 
D.\,R.\,Ward
 \\ \it
University of Cambridge, Cavendish Laboratory, J J Thomson Avenue, CB3 0HE, UK
}

\author{\centering 
D.\,Benchekroun, 
A.\,Hoummada, 
Y.\,Khoulaki
 \\ \it
Universit\'{e} Hassan II A\"{\i}n Chock, Facult\'{e} des sciences, B.P. 5366 Maarif, Casablanca, Morocco
}

\author{\centering 
J.\,Apostolakis, 
S.\,Arfaoui, 
M.\,Benoit, 
D.\,Dannheim, 
K.\,Elsener,
G.\,Folger, 
C.\,Grefe,
V.\,Ivantchenko, 
M.\,Killenberg$^b$, 
W.\,Klempt, 
   E.\,van der Kraaij, 
L.\,Linssen,
  A.\,-I.\,Lucaci-Timoce, 
A.\,M\"{u}nnich$^b$, 
  S.\,Poss, 
A.\,Ribon, 
P.\,Roloff, 
A.\,Sailer, 
D.\,Schlatter, 
E.\,Sicking, 
J.\,Strube$^c$,  
V.\,Uzhinskiy
 \\ \it 
CERN, 1211 Gen\`{e}ve 23, Switzerland
}

\author{\centering
C.\,C\^{a}rloganu, 
P.\,Gay, 
S.\,Manen, 
L.\,Royer
 \\ \it
Clermont Universit\'e, Universit\'e Blaise Pascal, CNRS/IN2P3, LPC, BP
10448, F-63000 Clermont-Ferrand, France
}

\author{\centering 
U.\,Cornett, 
D.\,David, 
A.\,Ebrahimi, 
G.\,Falley, 
N.\,Feege$^d$,
K.\,Gadow, 
P.\,G\"{o}ttlicher, 
C.\,G\"{u}nter,
O.\,Hartbrich, 
B.\,Hermberg, 
S.\,Karstensen, 
F.\,Krivan,
K.\,Kr\"uger, 
S.\,Lu$^e$,  
B.\,Lutz,
S.\,Morozov, 
V.\,Morgunov$^f$,  
C.\,Neub\"user,
M.\,Reinecke, 
F.\,Sefkow, 
P.\,Smirnov,
M.\,Terwort
 \\ \it
DESY, Notkestrasse 85,
D-22603 Hamburg, Germany
}

\author{\centering
A.\,Fagot,
M.\,Tytgat,
N.\,Zaganidis
 \\ \it
Ghent University, Department of Physics and Astronomy,
Proeftuinstraat 86, B-9000 Gent, Belgium
}

\author{\centering 
J.\,-Y.\,Hostachy, 
L.\,Morin
 \\ \it
Laboratoire de Physique Subatomique et de Cosmologie - Universit\'{e} Joseph Fourier Grenoble 1 -
CNRS/IN2P3 - Institut Polytechnique de Grenoble,
53, rue des Martyrs,
38026 Grenoble CEDEX, France
}

\author{\centering  
E.\,Garutti, 
S.\,Laurien, 
I.\,Marchesini$^g$, 
M.\,Matysek, 
M.\,Ramilli
 \\ \it
Univ. Hamburg,
Physics Department,
Institut f\"ur Experimentalphysik,
Luruper Chaussee 149,
D-22761 Hamburg, Germany
}

\author{\centering 
K.\,Briggl, 
P.\,Eckert, 
T.\,Harion, 
H.\,-Ch.\,Schultz-Coulon, 
W.\,Shen, 
R.\,Stamen
 \\ \it
 University of Heidelberg, Fakult\"at f\"ur Physik und Astronomie,
Albert \"Uberle Str. 3-5, 2.OG Ost,
D-69120 Heidelberg, Germany
}

\author{\centering 
S.\,Chang, A.\,Khan, D.\,H.\,Kim, D.\,J.\,Kong, Y.\,D.\,Oh
\\ \it
Department of Physics, Kyungpook National University, Daegu, 702-701,
Republic of Korea
}

\author{\centering 
B.\,Bilki$^h$, 
E.\,Norbeck$^a$, 
D.\,Northacker,
Y.\,Onel
 \\ \it
University of Iowa, Dept.\ of Physics and Astronomy,
203 Van Allen Hall, Iowa City, IA 52242-1479, USA
}

\author{\centering 
G.\,W.\,Wilson
 \\ \it
University of Kansas, Department of Physics and Astronomy,
Malott Hall, 1251 Wescoe Hall Drive, Lawrence, KS 66045-7582, USA
}

\author{\centering 
K.\,Kawagoe,
Y.\,Miyazaki, 
Y.\,Sudo,
H.\,Ueno, 
T.\,Yoshioka
 \\ \it
Department of Physics, Kyushu University, Fukuoka 812-8581, Japan
}

\author{\centering 
P.\,D.\,Dauncey
 \\ \it
Imperial College London, Blackett Laboratory,
Department of Physics,
Prince Consort Road,
London SW7 2AZ, UK 
}

\author{\centering 
E.\,Cortina Gil, 
S.\,Mannai
 \\ \it
Center for Cosmology, Particle Physics and Cosmology (CP3)
Universit\'{e} Catholique de Louvain, Chemin du cyclotron 2,
1320 Louvain-la-Neuve, Belgium
}

\author{\centering 
 G.\,Baulieu, 
 P.\,Calabria, 
 L.\,Caponetto, 
 C.\,Combaret, 
 R.\,Della\,Negra, 
 R.\,Et\'{e}, 
 G.\,Grenier, 
 R.\,Han, 
 J-C.\,Ianigro, 
 R.\,Kieffer, 
 I.\,Laktineh, 
 N.\,Lumb, 
 H.\,Mathez, 
 L.\,Mirabito, 
 A.\,Petrukhin, 
 A.\,Steen, 
 W.\,Tromeur, 
 M.\,Vander\,Donckt, 
 Y.\,Zoccarato 
  \\ \it
Universit\'{e} de Lyon, Universit\'{e} Lyon 1, 
CNRS/IN2P3, IPNL 4, rue E.\ Fermi, 69622
Villeurbanne CEDEX, France
}

\author{\centering 
J.\, Berenguer~Antequera, 
E.\,Calvo~Alamillo, 
M.-C.\, Fouz, 
J.\,Puerta-Pelayo 
 \\ \it
CIEMAT, Centro de Investigaciones Energeticas, Medioambientales y Tecnologicas, Madrid, Spain 
}

\author{\centering 
F.\,Corriveau
 \\ \it
Institute of Particle Physics of Canada and Department of Physics
Montr\'{e}al, Quebec,
Canada H3A 2T8
}

\author{\centering 
B.\,Bobchenko, 
M.\,Chadeeva, 
M.\,Danilov$^i$, 
A.\,Epifantsev, 
O.\,Markin, 
R.\,Mizuk$^i$, 
E.\,Novikov, 
V.\,Rusinov, 
E.\,Tarkovsky
 \\ \it
Institute of Theoretical and Experimental Physics, B. Cheremushkinskaya ul. 25,
RU-117218 Moscow, Russia
}

\author{\centering 
V.\,Kozlov, 
Y.\,Soloviev
 \\ \it
P.\,N.\, Lebedev Physical Institute,
Russian Academy of Sciences,
117924 GSP-1 Moscow, B-333, Russia
}

\author{\centering 
D.\,Besson, 
P.\,Buzhan,
A.\,Ilyin, 
V.\,Kantserov, 
V.\,Kaplin, 
E.\,Popova, 
V.\,Tikhomirov
\\\it
Moscow Physical Engineering Inst., MEPhI,
Dept.\ of Physics,
31, Kashirskoye shosse,
115409 Moscow, Russia
}

\author{\centering 
M.\,Gabriel, 
C.\,Kiesling, 
K.\,Seidel, 
F.\,Simon$^\spadesuit$, 
C.\,Soldner, 
M.\,Szalay, 
M.\,Tesar, 
L.\,Weuste 
 \\ \it
Max Planck Inst.\ f\"ur Physik,
F\"ohringer Ring 6,
D-80805 Munich, Germany
}

\author{\centering 
M.\,S.\,Amjad, 
J.\,Bonis, 
S.\,Conforti\,di\,Lorenzo, 
P.\,Cornebise, 
J.\,Fleury, 
T.\,Frisson, 
N.\,van der Kolk,  
F.\,Richard, 
R.\,P\"oschl, 
J.\,Rou\"en\'e
 \\ \it
Laboratoire de l'Acc\'{e}l\'{e}rateur Lin\'{e}aire, Centre
Scientifique d'Orsay, Universit\'{e} de Paris-Sud XI, CNRS/IN2P3, BP
34, B\^atiment 200, F-91898 Orsay CEDEX, France
}

\author{\centering 
 M.\,Anduze, 
 V.\,Balagura, 
 E.\,Becheva, 
 V.\,Boudry, 
 J-C.\,Brient, 
 R.\,Cornat, 
 M.\,Frotin, 
 F.\,Gastaldi,  
 E.\,Guliyev$^j$, 
 Y.\,Haddad, 
 F.\,Magniette, 
 M.\,Ruan$^k$, 
 T.H.\,Tran, 
 H.\,Videau
 \\ \it
 Laboratoire Leprince-Ringuet (LLR)  -- \'{E}cole Polytechnique, CNRS/IN2P3, F-91128 Palaiseau, France
}

\author{\centering 
S.\,Callier, 
F.\,Dulucq, 
G.\,Martin-Chassard, 
Ch.\,de la Taille, 
L.\,Raux, 
N.\,Seguin-Moreau
 \\ \it
OMEGA -- \'{E}cole Polytechnique, CNRS/IN2P3, F-91128 Palaiseau, France
}

\author{\centering 
J.\,Zacek 
 \\ \it
Charles University, Institute of Particle \& Nuclear Physics,
V Holesovickach 2,
CZ-18000 Prague 8, Czech Republic  
}

\author{\centering 
J.\,Cvach, 
P.\,Gallus, 
M.\,Havranek, 
M.\,Janata, 
J.\,Kvasnicka, 
D.\,Lednicky, 
M.\,Marcisovsky,  
I.\,Polak, 
J.\,Popule, 
L.\,Tomasek, 
M.\,Tomasek, 
P.\,Ruzicka, 
P.\,Sicho, 
J.\,Smolik, 
V.\,Vrba, 
J.\,Zalesak 
 \\ \it
Institute of Physics, Academy of Sciences of the Czech Republic, Na Slovance 2,
CZ-18221 Prague 8, Czech Republic
}

\author{\centering 
B.\,Belhorma, 
H.\,Ghazlane 
 \\ \it
Centre National de l'Energie, des Sciences et des Techniques Nucl\'{e}aires, 
B.P. 1382, R.P. 10001, Rabat, Morocco
}

\author{\centering              
K.\,Kotera, 
H.\,Ono, 
T.\,Takeshita, 
S.\,Uozumi
 \\ \it
Shinshu Univ.\,,
Dept. of Physics,
3-1-1 Asaki,
Matsumoto-shi, Nagano 390-861,
Japan
}

\author{\centering              
J.\,S.\,Chai, H.\,S.\,Song, S.\,H.\,Lee
 \\ \it
College of Information and Communcation Engineering / WCU Department
of Energy Science, Sungkyunkwan University, Suwon, 440-746, Repubic of
Korea 
}

\author{{\centering 
M.\, G\"otze, 
J.\,Sauer, 
S.\,Weber, 
C.\,Zeitnitz
 \\ \it
Bergische Universit\"{a}t Wuppertal,
Fachbereich C Physik,
Gaussstrasse 20,
D-42097 Wuppertal, Germany
}
 \\ \it
$^\spadesuit$ Corresponding author\newline
E-mail: \email{fsimon@mpp.mpg.de}
}

\author{  \\
\llap{$^a$}Deceased\\
\llap{$^b$}Now at DESY Hamburg, Germany\\
\llap{$^c$}Now at Tohoku University, Japan\\
\llap{$^d$}Now at Stony Brook University (SUNY), Dept.\ of Physics and
Astronomy, Stony Brook, NY, USA\\
\llap{$^e$}Now at Hamburg University, Germany \\
\llap{$^f$}On leave from ITEP\\
\llap{$^g$}Also at DESY Hamburg, Germany\\
\llap{$^h$}Also at Argonne National Laboratory\\
\llap{$^i$}Also at MEPhI and at Moscow Institute of Physics and Technology\\
\llap{$^j$}TRIUMF, Vancouver, BC, Canada\\
\llap{$^k$}Now at IHEP, Beijing, China

}

\abstract{The intrinsic time structure of hadronic showers influences the timing capability and the required integration time of hadronic calorimeters in particle physics experiments, and depends on the active medium and on the absorber of the calorimeter. With the CALICE T3B experiment, a setup of 15 small plastic scintillator tiles read out with Silicon Photomultipliers, the time structure of showers is measured on a statistical basis with high spatial and temporal resolution in sampling calorimeters with tungsten and steel absorbers. The results are compared to \geant\ (version 9.4 patch 03) simulations with different hadronic physics models. These comparisons demonstrate the importance of using high precision treatment of low-energy neutrons for tungsten absorbers, while an overall good agreement between data and simulations for all considered models is observed for steel.}

\begin{document}

\bibliographystyle{JHEP} 

\section{Introduction}
\label{sec:Introduction}

Hadronic calorimeters are key systems in modern collider experiments, contributing substantially to the measurement of jets and missing energy, and often also to triggering and background suppression based on timing. For the latter, an understanding of the intrinsic time structure of hadronic showers is important. This is particularly relevant at colliders with high background rates and short time spacing between collisions, such as the future multi-TeV Compact Linear Collider (CLIC) \cite{Lebrun:2012hj}, with a bunch crossing rate of 2 GHz and high beam-induced background rates from $\gamma\gamma \rightarrow$ hadrons processes. 

The time structure of hadronic showers is characterized by prompt energy depositions by relativistic hadrons and by electrons and positrons from electromagnetic subshowers following $\pi^0$ and $\eta$ production in the cascade, and by a delayed component extending to $\mu$s time scales, mainly from neutron-induced processes such as elastic scattering and particle emission following neutron capture. The relative importance of these components depends on the absorber material, which strongly influences the amount of produced neutrons, and on the active material, where in particular the hydrogen content determines the direct sensitivity to MeV-scale neutrons. 

For the barrel hadron calorimeters at CLIC, tungsten is considered as absorber material to achieve a maximum compactness of the detector while also obtaining good containment for jets with TeV energies \cite{Linssen:2012hp}. To study the behavior of hadron calorimeters with tungsten absorbers and to validate the simulation of such a calorimeter, the CALICE collaboration has constructed and tested a $\sim$ 1 m$^3$ prototype of a tungsten calorimeter with highly granular scintillator readout (W-AHCAL) \cite{Adloff:2013jqa}, based on the active elements of the CALICE analog hadron calorimeter (AHCAL) \cite{Adloff:2010hb}. In the context of CLIC, the timing aspects of such a calorimeter are of particular importance, since they play a key role in the rejection of beam-induced hadronic background \cite{Marshall:2012ry}. Since the AHCAL elements do not provide a time resolved readout, the Tungsten Timing Test-Beam (T3B) experiment \cite{Simon:2013zya} was included in the test beam campaigns to obtain a first measurement of the time structure of hadronic showers. T3B is an add-on detector to the CALICE calorimeters with 15 small scintillator cells read out with sub-ns time resolution over a time window of 2.4 $\mu$s, and can provide information on the time structure on a statistical basis by studying large data samples. To provide data for steel absorbers in addition to the tungsten data, the T3B experiment also took data together with the CALICE semi-digital hadron calorimeter (SDHCAL) \cite{Laktineh:2011zz}, which is based on resistive plate chamber (RPC) readout and a steel absorber structure. Since the optimization of the detector concepts for future colliders and the evaluation of their performance is based on \geant\ simulations, the validation of the modelling of the time structure of the calorimeter response is of high importance. Thus, simulations performed with different hadronic shower models are confronted with the data taken with the T3B setup. 

This article presents the results of the T3B experiment obtained during two test beam periods at the CERN SPS, one in September 2011 together with the W-AHCAL and one in October 2011 together with the SDHCAL. A brief introduction to the physics of hadronic showers is given in Section \ref{sec:PhysicsIntro} and the test beam setup is described in Section \ref{sec:Setup}, followed by a short description of the data reconstruction and of the simulation in Sections  \ref{sec:CalibReco} and \ref{sec:Sim}. The results and the comparisons of data and simulations are presented in Section \ref{sec:StandAnalysis}.


\section{The Time Structure of Hadronic Showers}
\label{sec:PhysicsIntro}

The variety of interactions which take place within hadronic cascades results in a complex structure of hadronic showers, which also has consequences for the time evolution of the signal of calorimeters. A detailed discussion of the underlying physical processes can be found in \cite{Wigmans} and references therein. 

The cascade starts with the inelastic interaction of the incoming highly energetic hadron, which produces secondary relativistic hadrons as well as spallation nucleons with typical energies of 100 MeV and fragments from the destruction of the nucleus the hadron interacted with. The excited nuclear remnants emit evaporation nucleons, predominantly neutrons with energies in the MeV range. The energetic particles induce further nuclear reactions, leading to the creation of a hadronic shower, while low-energy charged particles get stopped in the material. The neutrons, with a substantially larger mean free path, spread throughout the calorimeter and get moderated by elastic and inelastic interactions, and result in neutron capture processes when reaching eV-scale energies. While the creation of relativistic hadrons and the spallation processes occur quasi-instantaneously, the evaporation processes extend to ns time-scales and the neutron captures occur substantially delayed due to the flight time of the neutrons during the moderation process, extending to time scales of $\mu$s. 
 
In the calorimeter, the energy depositions from relativistic hadrons as well as from the electromagnetic part of the shower originating from the decay of $\pi^0$s and $\eta^0$s created in  highly energetic inelastic reactions result in a prompt signal component. The evaporation neutrons provide signals on the few to few tens of ns, predominantly by elastic interactions with the active detector medium. For hydrogenous active components, such as plastic scintillators, this signal component is thus expected to be enhanced with respect to detectors with low hydrogen content. The neutron capture processes also result in detectable signals if they occur close to the interface between absorbers and active elements, leading to a very late signal component extending to $\mu$s time scales. Since the late signals are predominantly due to neutrons which spread out in the calorimeter while the relativistic particles are concentrated along the shower core, their relative importance is expected to increase with increasing distance from the shower axis. 

The absorber material has a substantial influence on the relative importance of the late components of the cascade. The amount of neutrons produced in spallation and evaporation processes depends on the nucleus, and increases for heavier nuclei. In lead, for example, the number of neutrons is estimated to be approximately a factor of 3 to 4 higher than in steel \cite{Wigmans}, with the largest enhancement seen for evaporation neutrons. For tungsten, a similar (but somewhat smaller) enhancement with respect to steel is expected, based on the size of the nucleus. Experimentally, a larger delayed signal component in comparison to steel absorbers has been observed in uranium calorimeters \cite{DeVincenzi:1985fz}, where fission processes provide an additional neutron source. Detailed spatially unresolved measurements of the time structure in a uranium-scintillator sampling calorimeter show contributions from recoiling protons on the few 10 ns time scale, and signals from photons following neutron capture on the few 100 ns to $\mu$s time scale \cite{Caldwell:1992te}, consistent with the picture of the hadronic shower evolution discussed above. For a scintillating fiber / lead calorimeter, an exponentially decaying late component with a time constant of around 9.5 ns has been observed \cite{Acosta:1990bq}, demonstrating the importance of late shower components and correspondingly longer integration times also for lead-based calorimeters. In a copper-based dual readout calorimeter with plastic scintillator and quartz fibers a time constant of around 20 ns was observed \cite{Akchurin:2007uf}. The absence of this late signal in the Cherenkov component of the detector signal points to shower neutrons as the origin of this signal component. The measurements also show the expected increasing importance of the late shower component with increasing distance from the shower axis.

\section{The Experimental Setup}
\label{sec:Setup}

\begin{figure}
\begin{center}
  \includegraphics[width=.99\linewidth]{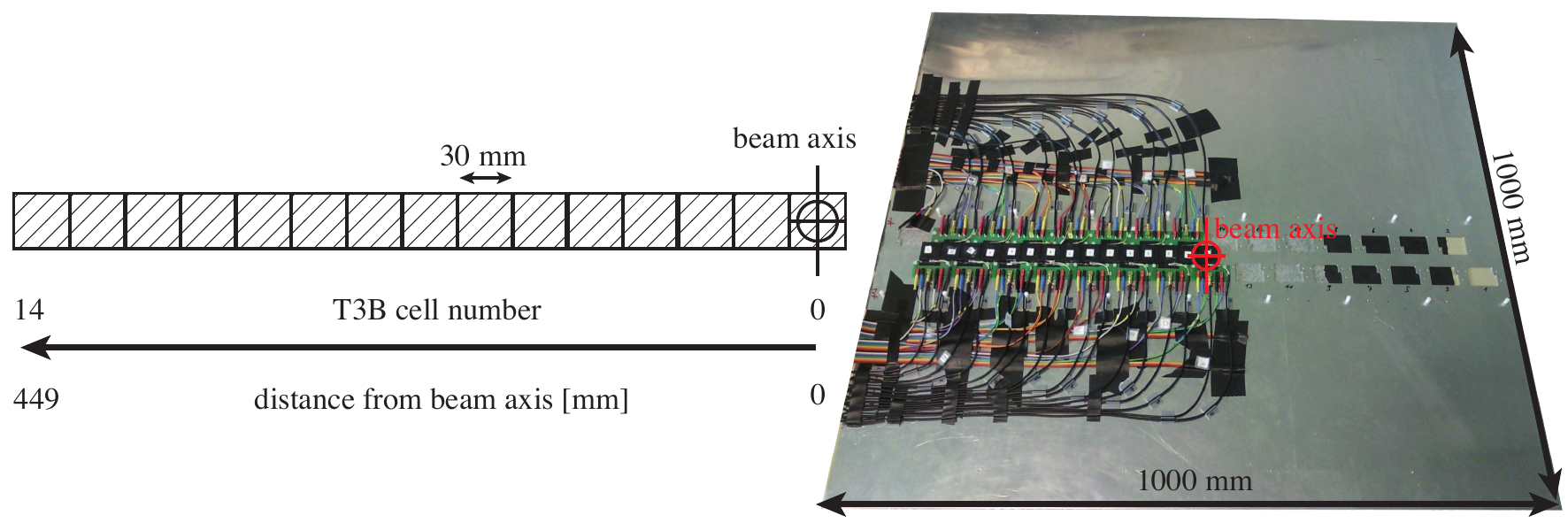}
  \caption{Illustration of the layout of the T3B detector (left), with 15 square scintillator tiles arranged in a strip from the beam axis outwards. The photograph (right) shows the opened T3B layer, with the scintillator cells connected to preamplifiers with SiPMs and the required cabling. The position of the beam axis is indicated.}
  \label{fig:Setup:Layer}
\end{center}
\end{figure}

\begin{figure}
\begin{center}
  \includegraphics[width=.99\linewidth]{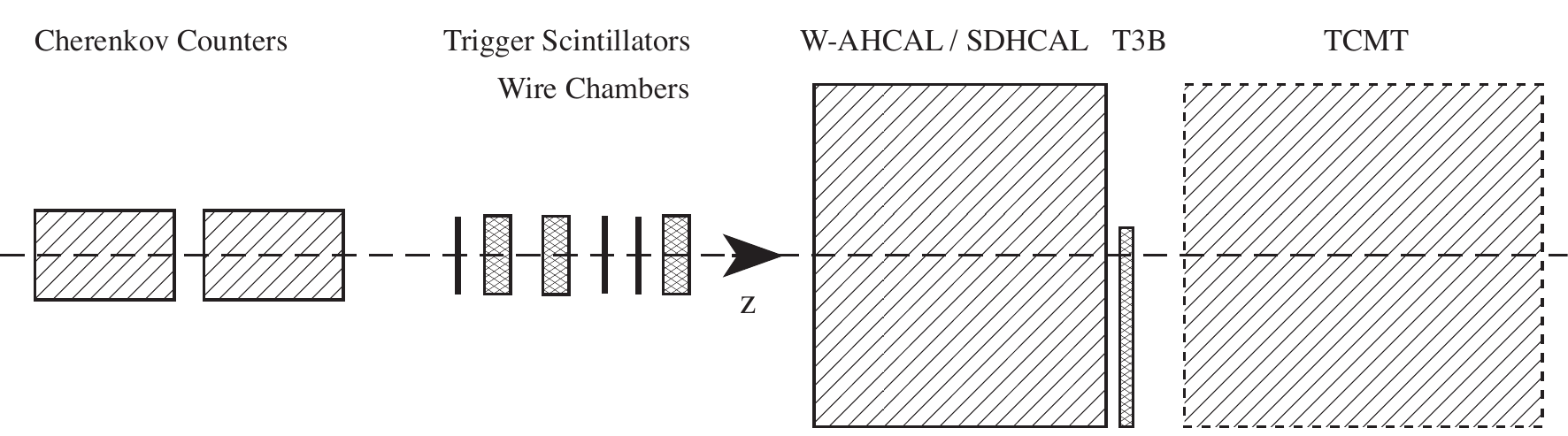}
  \caption{The setup at the test beam at the CERN SPS H8 beam line in the North Hall. The tail catcher TCMT was only present for runs with the CALICE W-AHCAL, and not installed for runs with the Fe-SDHCAL. Illustration not to scale.}
  \label{fig:Setup:Testbeam}
\end{center}
\end{figure}

\begin{table}[tp]

\caption{The components and their respective thickness $d$ of the layers of the three different detectors (T3B left, W-AHCAL center, SDHCAL right), listed by their appearance when following the beam downstream. The plastic scintillator material is polystyrene in the case of the W-AHCAL and polyvinyltoluene in the case of T3B.}

\begin{center}

	\begin{minipage}[t]{.3\linewidth}
		\begin{tabular}[t]{l|c} 
			\multicolumn{2}{c}{T3B Layer} \\
			\hline
			Component & $d$ [mm]\\	 
			\hline
			\hline
			Al Cassette & 1.0 \\ 
			Air & 2.3 \\ 
			Scintillator & 5.0 \\ 
			Air & 1.0 \\ 
			PCB & 1.7 \\ 
			Al Cassette & 2.0 \\ 
			\hline
			Total & 13 \\ 	
		\end{tabular}
	\end{minipage}
	\hspace{.01\linewidth}
	\begin{minipage}[t]{.3\linewidth}
		\begin{tabular}[t]{l|c} 
			\multicolumn{2}{c}{W-AHCAL Layer} \\
			\hline
			Component & $d$ [mm]\\	 
			\hline
			\hline
			Steel Support & 0.5 \\ 
			Tungsten & 10 \\ 
			Air & 1.25 \\ 
			Steel Cassette & 2.0 \\ 
			Cable Mix & 1.5 \\ 
			PCB & 1.0 \\ 
			Scintillator & 5.0 \\ 
			Steel Cassette & 2.0 \\ 
			Air & 1.25 \\ 
			\hline
			Total & 24.5 \\ 			
		\end{tabular}
	\end{minipage}
	\hspace{.01\linewidth}
	\begin{minipage}[t]{.3\linewidth}
		\begin{tabular}[t]{l|c} 
			\multicolumn{2}{c}{Fe-SDHCAL Layer} \\
			\hline
			Component & $d$ [mm]\\	 
			\hline
			\hline
			Steel 304L & 20 \\ 
			Epoxy & 1.6 \\ 
			PCB  & 1.2 \\ 
			Mylar & 0.23 \\ 
			Graphite & 0.1 \\ 
			Glass & 1.8 \\ 
			RPC Gas & 1.2 \\ 
			\hline
			Total & 26.13 \\ 			
		\end{tabular}
	\end{minipage}

\end{center}
\label{tab:Setup:MaterialBudget}
\end{table}

The T3B experimental setup consists of one strip of 15  plastic scintillator tiles \cite{Simon:2010hf} with a size of $30\times 30 \times 5$ mm$^3$, read out by Hamamatsu MPPC silicon photomultipliers (SiPMs), as illustrated in Figure \ref{fig:Setup:Layer}. The first scintillator tile, labeled "0" in the figure, is centered on the beam axis, so that T3B  samples the shower from the central core out to a radius of 449 mm. The complete setup is encased in an aluminium cassette with additional temperature sensors to monitor the environmental conditions. Each photon sensor is read out via a preamplifier board connected to one of the inputs of 4-channel USB oscilloscopes\footnote{Picotech PicoScope 6403} with a sampling frequency of 1.25 GS/s, 8 bit vertical resolution and a readout window of 2.4 $\mu$s per trigger. More details on the T3B setup are given in  \cite{Simon:2013zya}. The material budget of the T3B layer along the beam axis is shown in Table \ref{tab:Setup:MaterialBudget}, \emph{left}.

In the test beam experiments, the T3B layer was placed directly behind the respective hadron calorimeter prototype (W-AHCAL or SDHCAL), in an arrangement as shown in Figure  \ref{fig:Setup:Testbeam}. For the data taking period together with the W-AHCAL, a tail catcher and muon tracker (TCMT) with steel absorbers was installed 90 mm downstream of T3B. This additional detector was not present for data taking with the SDHCAL. Simulation studies with and without this additional detector have shown that the presence of the TCMT does not have a noticeable effect on the results discussed below. Data from this detector, as well as data from the wire chambers upstream of the calorimeters is not used in the present analysis since it was not available for data taking with the SDHCAL. 

The W-AHCAL consists of 38 instrumented layers, each with a material composition as given in Table \ref{tab:Setup:MaterialBudget}, \emph{center}. The tungsten used for the absorber layers is an alloy with a density of 17.8 g/cm$^3$ consisting of 92.99\% W, 5.25\% Ni and 1.76\% Cu. The lateral size of the calorimeter is $0.99 \times 0.99$ m$^2$, and the depth is 931 mm, corresponding to 5.1 $\lambda_I$. The tungsten in each layer is arranged in an octagonal shape embedded in a square aluminium frame, with a lateral extension of the tungsten of 0.81 m in the region where T3B was installed. The absorber elements consist of 10 mm thick tungsten plates and a 0.5 mm thick steel support layer . More details on the W-AHCAL are given in \cite{Adloff:2013jqa}. T3B was installed in the mechanical support structure of the W-AHCAL where layer 40 would be, with 34.5 mm of air separating T3B from the readout cassette of the last calorimeter layer. The mechanical support structure of the calorimeter includes an aluminium frame which partially covers the outermost T3B tile. Consequently, this tile is not used in the data analysis to avoid a bias from the different material composition in front of that tile. 

The SDHCAL consists of a total of 50 absorber layers of machined stainless steel (type 304L) with a thickness of 20 mm, with the first 40 layers equipped with RPC chambers for readout, and the last ten layers left empty in the test beam period considered here. The lateral size of the detector is $1\times 1$ m$^2$, and the depth is 1306.5 mm, corresponding to 6.5 $\lambda_I$. The material composition of one SDHCAL layer equipped with an RPC is given in Table \ref{tab:Setup:MaterialBudget}, \emph{right}. Here, T3B was installed directly behind the last absorber layer, with no air gap separating T3B from the last SDHCAL absorber layer. 

For both setups the event recording of T3B was triggered by a coincidence of two scintillator sensors upstream of the respective calorimeter prototype. When operating together with the W-AHCAL, the trigger signal was taken from the data acquisition system of the W-AHCAL, providing the possibility for a synchronized analysis of T3B data and data from the main calorimeter. During data taking together with the SDHCAL, the trigger was generated by a coincidence unit directly from the trigger scintillator signals, since the SDHCAL data acquisition was still being commissioned in the beam period used, providing insufficient trigger rate to achieve the large event numbers required by T3B. An additional gate generator was used to veto additional triggers for 3 $\mu$s after an accepted trigger to allow the T3B DAQ to record the full time window for each event and to provide sufficient time for all oscilloscopes to return to the state of waiting for triggers. Since the trigger rate was typically at the level of a few 100 Hz this did not result in a significant dead time of the data acquisition. For both setups, the trigger scintillator coincidence signal was also recorded on one channel of the T3B data acquisition, to identify potential double particle events and to reject calibration triggers of the W-AHCAL DAQ. 

Particle identification was possible via two threshold Cherenkov detectors upstream of the calorimeter prototypes, with their thresholds set to distinguish between pions and kaons and between kaons and protons. The signals of these Cherenkov detectors were also recorded by the T3B DAQ, providing the potential for particle identification, which is however not used in the analysis described in the present paper. Further details on the T3B DAQ are given in \cite{Simon:2013zya}.

\section{Data Set and Event Reconstruction}
\label{sec:CalibReco}

The data samples considered in this paper are taken from run periods in September 2011 together with the W-AHCAL, and in October 2011 with the SDHCAL. In both test beam programs data were collected at various energies and with both positive and negative polarity. Due to the small active area of the T3B experiment, large event samples are needed for a precise measurement of the late shower components, so we concentrate here on the analysis of data taken with positive mixed hadron beams at 60 GeV, which is by far the largest data sample available at one energy for both configurations. For the runs with the W-AHCAL and the SDHCAL the same beam settings were used, resulting in an identical muon contamination of the hadron beams of approximately 5\%. Since no tail catcher that could be used as a muon veto was available for the SDHCAL data run, and a full synchronisation of the data streams of the W-AHCAL and T3B was not performed, muons events could not be rejected. They are present in both the tungsten absorber and the steel absorber data set. 
In addition to the hadron runs, large samples of muon data with an energy of \mbox{180 GeV} were collected for calibration purposes together with the SDHCAL.   

The data are stored as raw waveforms by the T3B DAQ. These waveforms are further processed by a dedicated calibration and reconstruction software, which identifies the time of each firing microcell of the photon sensor and provides a time distribution of identified photon equivalents (p.e.) with sub-nanosecond resolution. From the identified single photon signals, detector hits are reconstructed by requiring a minimum of 8 p.e.\ within a time window of 9.6 ns. This time window is chosen since it approximately corresponds to the recovery time of the photon sensor, thus making the occurrence of afterpulses within this time window very unlikely. These pulses originate from the delayed release of trapped electrons in the SiPM, resulting in additional pulses which are correlated with the original signal. Since the the probability for additional cell breakdowns are reduced during the recharging period of a cell, afterpulses are unlikely within the first few ns after the original signal. The hits are calibrated in energy using the scale given by the most probable signal of a minimum ionizing particle traversing a T3B tile, 1 MIP, which corresponds to a visible energy deposit of 805 keV in the scintillator, as determined from \geant~simulations. The reconstruction threshold of 8 p.e.\  corresponds to approximately 0.4 MIP. The timing of reconstructed hits is given by the time of the second p.e., to mitigate the influence of the thermal dark rate of the photon sensors. This procedure introduces an amplitude-dependent bias of the reconstructed hit time, which is corrected for with a time-slewing correction. For the analysis presented in this paper, only the first hit in each detector channel in a given event is considered. The time of this hit is referred to as time of first hit (TofH) in the following. The use of only the first hit in each detector channel avoids the influence of afterpulsing of the photon sensor.  At the same time, the impact of the restriction to the first hit in each detector cell is quite small, since the high granularity of the readout makes multiple hits of the same cell at different times during one event unlikely. Dedicated studies have shown that the probability for multiple hits separated by more than 10 ns is at the few percent level, with a conservative upper limit of 12\%. This limit is largely influenced by the high rate of afterpulsing. Further details of the calibration and reconstruction procedure are given in \cite{Simon:2013zya}. 

A time calibration is performed to determine the relative time offsets between the different data sets due to the use of different trigger systems and other changes in the experimental setup which influence signal run times. This calibration uses the distribution of all identified first hits of the central T3B tile only, to avoid possible time-of-flight effects and influences of the different beam profiles in muon and hadron runs. The time resolution of the complete system, which is influenced  by the intrinsic time resolution of the T3B scintillator cells including the readout system, by the data analysis procedure and by the time jitter of the trigger, is extracted from the width of the main peak of the timing distribution for each data set. The width of the distribution is determined by a Gaussian fit. Table \ref{tab:TimeResolution} gives the available statistics for each data set, the events with at least one T3B hit which are further analyzed, the overall number of identified first hits and the resulting average multiplicity in T3B as well as the determined timing precision, which ranges from 1.07 ns to 0.7 ns for muons in steel and for pions in tungsten, respectively. The differences in precision are due to the different trigger setups used. The highest precision is obtained for hadron data taken with the W-AHCAL, where the CALICE trigger system with a coincidence of two $10 \times 10 \times 1$ cm$^3$ trigger scintillators with PMT readout is used. For the steel absorber data, taken with the SDHCAL, a stand-alone trigger was used for both muon and hadron runs. This system has a larger time jitter due to the use of different scintillators, PMTs and different coincidence electronics. For data taking with muons, larger trigger scintillators are used to increase the area of the main calorimeter that is covered by the trigger, making use of the wide profile of the muon beam. These larger scintillators result in an additional time jitter, making the muon data set the one with the lowest timing precision. 

\begin{table}[t]
\caption{Timing precision and acquired statistics of the different data sets. The muon sample was acquired with steel absorber.}
\begin{center}
	\begin{tabular}{l|ccc}
		Data Set & $180\,\text{GeV}$ muons& $60\,\text{GeV}$ hadrons - steel & $60\,\text{GeV}$ hadrons - tungsten\\ \hline \hline
		Timing Precision & $1.07\,\text{ns}$ & $1.00\,\text{ns}$ & $0.70\,\text{ns}$ \\
		Events in Analysis & 5.40M & 1.60M & 4.06M \\ 
		$\#$ Events with Hits & 790k & 203k & 716k \\ 
		$\#$ First Hits & 854k & 312k & 1014k \\
		Multiplicity & 1.08 & 1.54 & 1.42 \\ \hline
	\end{tabular}
\end{center}
\label{tab:TimeResolution}
\end{table}


\section{Simulations}
\label{sec:Sim}
In addition to providing precise information on the average time structure of hadronic showers in different absorber materials, a goal of the T3B experiment is to validate hadronic models used for Monte Carlo based simulations of hadronic cascades provided by the \geant~toolkit\cite{Agostinelli:2002hh,Geant4:PhysicsLists}. The simulations were performed using \geant~version 9.4p03, with the full experimental setup of the main calorimeter and T3B implemented using the material composition listed in Table \ref{tab:Setup:MaterialBudget}.  Saturation effects in the scintillator for heavily ionizing particles are modelled using Birk's law \cite{Birks1964} to account for the reduced visible energy for the interactions of low-energy recoil protons and nuclear fragments. For this, the standard implementation in \geant~is used, provided by the class \texttt{G4EmSaturation}. This parametrizes the saturation effect as $\frac{dL}{dx} \propto \frac{dE}{dx}/(1+ k_B \cdot \frac{dE}{dx})$, where $\frac{dL}{dx}$ is the light output per unit length, $\frac{dE}{dx}$ the energy deposition by the particle, and $k_B$ a material-dependent parameter describing the saturation. In the present study, the \geant~default saturation parameter for plastic scintillator is used, given by $k_B$ = 0.0794 mm/MeV \cite{Hirschberg:1992xd}. In addition to the simulation of the interaction of particles in the detector provided by \geant, the T3B detector response is simulated by a data-driven digitization procedure \cite{Simon:2013zya} which accounts for the time distribution of photon signals originating from instantaneous energy depositions and for effects of photon statistics originating from the overall small numbers of detected photons.

\subsection{Hadronic Cascade Models}
\geant~implements various models for hadronic cascades, each covering a specific energy region. To provide a description of hadronic interactions over the full relevant energy range, several such models are grouped together into hadronic physics lists \cite{Geant4:PhysicsLists,Geant4:PhysicsLists:Improvements}. In this paper, three physics lists are used:  QGSP\_BERT, QGSP\_BERT\_HP and QBBC. For each physics list and each detector configuration 2 million pion events were simulated. 

QGSP\_BERT uses the quark-gluon-string model QGS \cite{Folger:2003sb}, together with a precompound model to handle the de-excitation of states created in the inelastic interaction, at energies above 12 GeV. The low energy parametrized LEP model is used at energies between 9.5 GeV and 25 GeV, and the Bertini cascade BERT \cite{Heikkinen:2003sc} at energies below 10 GeV. In the overlap of energy ranges between models a specific model is chosen for the simulation of each interaction, with a probability which is zero at the start (or end) of its applicability range and one at the energy at which it becomes the only applicable model. QGSP\_BERT\_HP is identical to the first list, except that it incorporates an additional high-precision (HP) extension for the realistic simulation of low energy neutrons below 20 MeV using measured cross sections. 

The third physics list considered is QBBC, which uses the QGS model together with a precompound model at energies above 12 GeV, the Fritiof string-parton model with a precompound model at energies between 4 GeV and  25 GeV and the Bertini cascade at energies below 5 GeV. Below 1.5 GeV the Binary cascade model is used for nucleons. Instead of the HP extension, it uses a different, computationally faster dedicated implementation for the tracking of slow neutrons. 

The selected physics lists provide the possibility for a study of the importance of the realism of the simulation of the slow neutron component and  allows a comparison of different implementations of neutron tracking.

\section{Results}
\label{sec:StandAnalysis}

In this section, the analysis results of a reconstructed and calibrated subset of the test beam data acquired with the T3B experiment are
presented. The analyzed data comprise hadron shower events (from impinging $\pi^+$ and protons, with a fractional contribution of 60\% and 40\%) at $60\,\text{GeV}$ as well as muon data at
$180\,\text{GeV}$ beam energy. Since muons do not induce hadronic cascades, their energy depositions occur promptly. At the same time, the muon data
are also subject to all detector effects. Thus, the timing of the hadron data sets is investigated relative to the standard signal of muons. Unless
stated otherwise, the analysis includes all T3B first hits in a time window of $-20\,\text{ns}$ to $200\,\text{ns}$ around the hardware
trigger time ($t$ = 0). Each hit is characterized by its time, which is determined with a precision of approximately 1 ns \cite{Simon:2013zya}, by its energy in units of MIP and by its distance from the beam axis, given by the position of the cell it occurs in. 

\begin{figure}[t]
		\centering
       \includegraphics[width=0.49\textwidth]{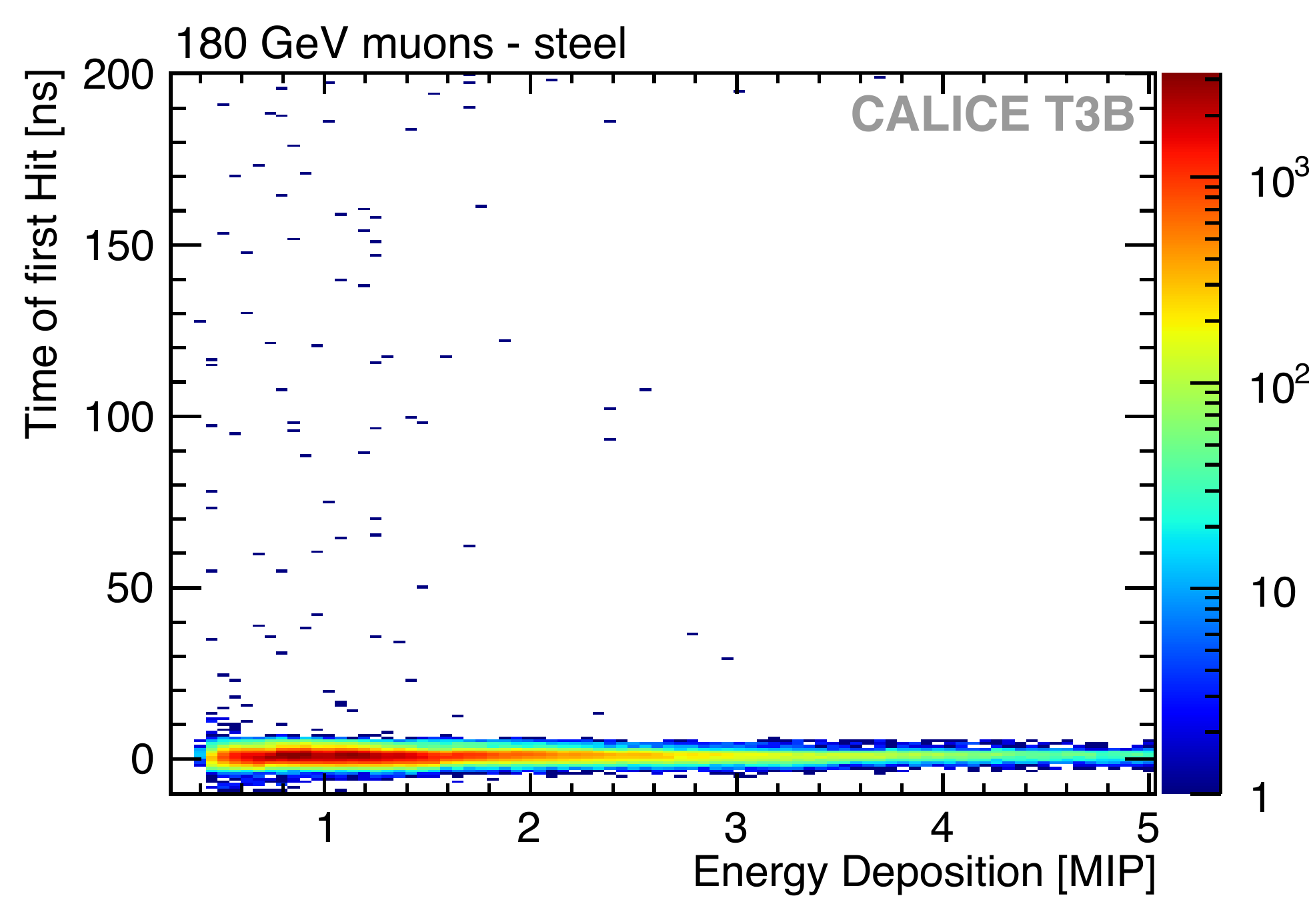} \\
       \includegraphics[width=0.49\textwidth]{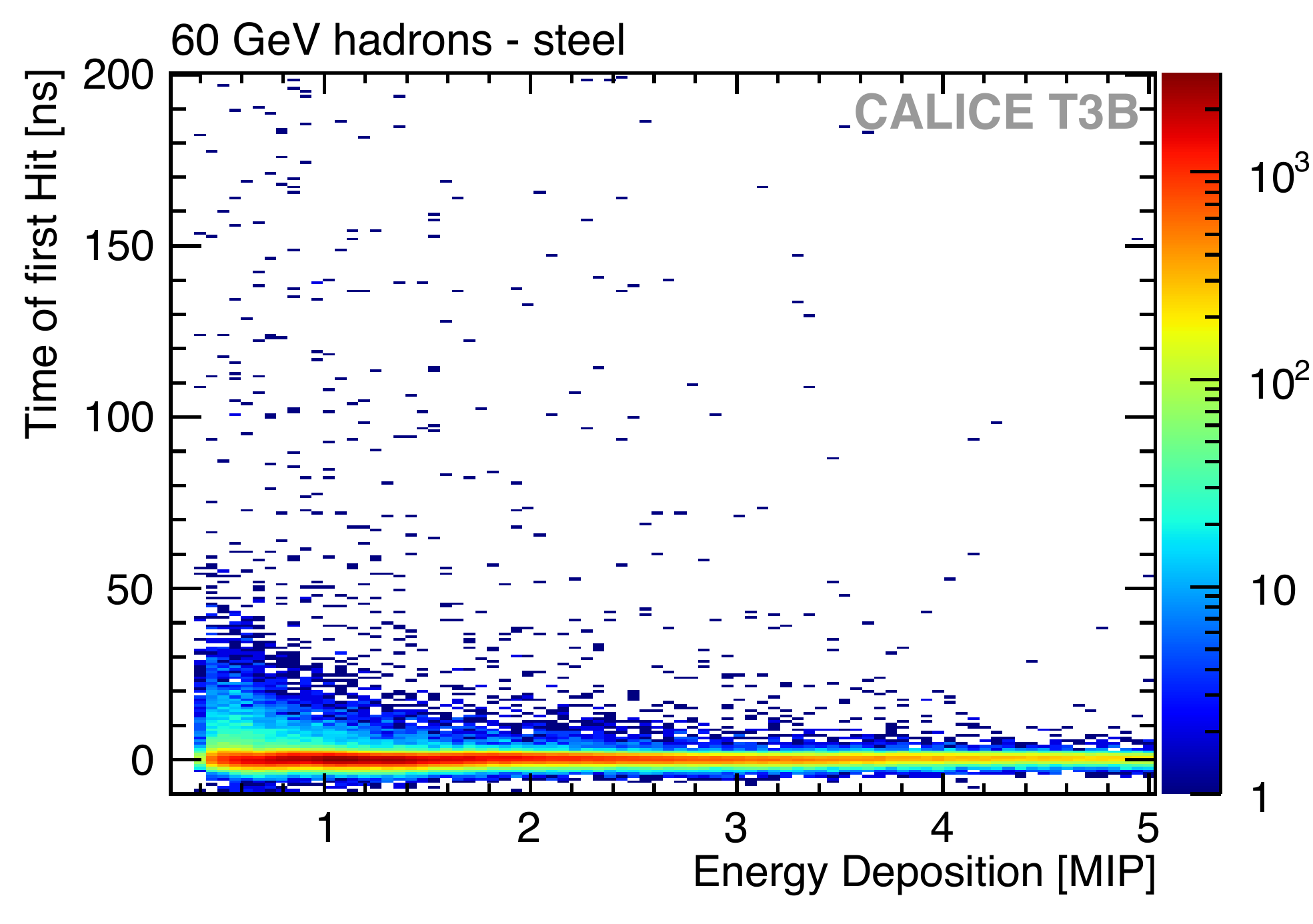}
       \includegraphics[width=0.49\textwidth]{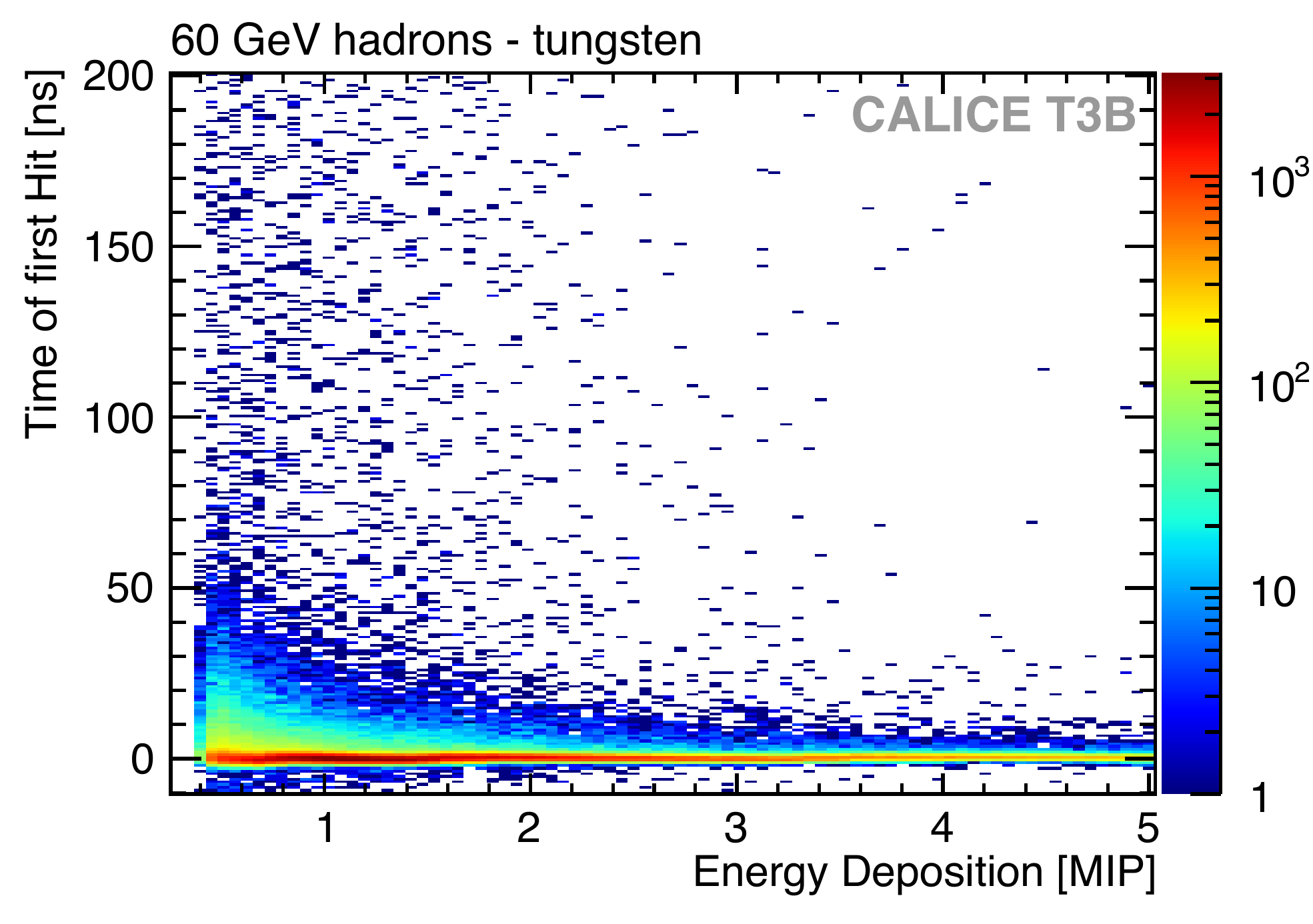}
        \caption{Distribution of $203103$ T3B events with identified first hits with respect to their time of occurrence and the energy deposited by them for different run characteristics, namely muon data (top) and hadron data in steel (bottom, left) and tungsten (bottom, right). The color-coded axis shows the number of entries per bin.}
        \label{fig:T3BAnalysis:TofH:2DDistribution}
\end{figure}

Figure \ref{fig:T3BAnalysis:TofH:2DDistribution} shows the distribution of hits for two out of the three variables, namely time and energy, for muons with steel absorbers as well as hadrons for both absorber materials. The hits in all T3B cells are combined, and the same number of events with at least one T3B hit is shown for all data sets to ease the comparison. The number of included events is defined by the hadron data set in steel, the data set with the lowest number of events with identified T3B hits, as shown in Table \ref{tab:TimeResolution}. These distributions already give a first impression of the differences between muons and hadrons, and the differences between the temporal shower development in steel and tungsten. The muon data are characterized by instantaneous energy depositions, with a very low additional contribution of low-energy late hits originating from thermal noise of the photon sensors. For hadrons a substantial late activity, in particular at lower hit energy, is visible. These delayed signals are more pronounced for tungsten than for steel. Projections of these distributions, with an additional dimension given by the T3B cell number, form the basis of the following more detailed analysis.

\subsection{Systematic Uncertainties}
\label{sec:Systematics}

For an assessment of the significance of the differences observed between data and simulations a full evaluation of possible systematic uncertainties is necessary. The time of first hit used in the present analysis is a very robust variable, which can be also very well described in simulations. Still, six possible sources of systematic uncertainties were identified:
\begin{itemize}
\item The relative timing between data sets: For each data set, both in data and in fully digitized simulations, the time offset which is used to define $t$ = 0 is determined with a Gaussian fit to the prompt signal component, as discussed further in \cite{Simon:2013zya}. The uncertainty of this determination is at the level of 100 ps to 200 ps, resulting in the assignment of a 0.2 ns systematic timing uncertainty for comparisons of different data sets.
\item The T3B position relative to the beam axis: T3B is installed in the W-AHCAL and in the SDHCAL such that the nominal beam axis goes through the center of tile 0. The uncertainties on the positioning of the T3B layer and of the beam is at the centimeter level. Conservatively, a systematic uncertainty of the size of one tile (3 cm) is assigned to the radial distance for the studies as a function of radius. This uncertainty is converted into a corresponding uncertainty in time using the radial dependence of the mean time of first hit of simulations with the QGSP\_BERT\_HP physics list, discussed in detail in Section \ref{sec:MC}. This uncertainty ranges from 0.2 ns at $r$ = 0 in both steel and tungsten to 0.4 ns and 0.7 ns at $r$ = 40.2 cm for steel and tungsten, respectively.
\item The T3B energy scale: The energy scale of each tile is determined in test bench measurements with a $^{90}$Sr source. The difference between the response to $^{90}$Sr electrons and MIPs is determined with muons from dedicated muon runs in the central T3B cell, and accounted for in the calibration. The uncertainty of the calibration is at the 1\% level, based on the uncertainty of the fits of the most probable value of the signals. In addition, corrections are made based on the measured gain of the photon sensor, as discussed in \cite{Simon:2013zya}. The precision of the gain determination is also at the 1\% level, leading to the assignment of an energy scale systematic of 2\% for the studies as a function of deposited energy. As for the case of the position uncertainty, this is converted to a corresponding uncertainty in time using the energy dependence of the mean time of first hit of simulations with the QGSP\_BERT\_HP physics list. At 0.5 MIP in tungsten, this results in an uncertainty of 0.2 ns, in steel the corresponding uncertainty is 0.1 ns. For hit energies above 0.8 MIP the uncertainty is below 0.02 ns for both tungsten and steel, and thus negligible.
\item The time-slewing correction: The correction for the amplitude dependence of the hit time determination may introduce a residual energy dependence of the hit time determination. Form the small residual slope of the time distribution for muon hits as a function of hit energy observed after the correction, the systematic uncertainty is estimated to be at the level of 0.1 ns. 
\item The geometrical difference between the tungsten and steel setups: There is an additional potential uncertainty when directly comparing tungsten and steel data due to the different depth of the location of T3B behind the W-AHCAL and SDHCAL, both geometrically and in units of interaction length. The possible impact of this was studied by performing simulations with different detector depths. These studies show a negligible uncertainty at small shower radii, and uncertainties smaller than 0.25 ns at the largest radii. This uncertainty affects only the direct comparison of the radial dependence of the mean time of first hit in tungsten and steel.
\item The normalization of the different data sets: When comparing the number of T3B hits per time bin observed in data and simulations, there is a normalization uncertainty related the number of true hadron events in the sample. While the number is perfectly well known for simulations, in data an uncertainty originates from the muon component of the beam and from detector noise which can also result in the fulfillment of the acceptance requirement of an event with a T3B hit. A 10\% uncertainty on the data normalization is assigned to account for this uncertainty when comparing the absolute number of hits per time bin. For bins where this value is smaller than the noise contributions observed in muon events the noise level is taken as the uncertainty instead.
\end{itemize}

From these individual uncertainties, the total uncertainties are obtained by adding the relevant ones for each measurement in quadrature. The resulting total systematic uncertainties for the data-MC comparisons for the distributions as a function of hit energy are from 0.3 ns to 0.2 ns in tungsten, and 0.2 ns for all energies in steel. 
For the distributions as a function of radial distance from the beam axis, the total systematic uncertainties range from 0.3 ns to 0.5 ns for steel and from 0.3 ns to 0.7 ns for tungsten. For the comparison of the radial dependence of the mean time of first hit observed in data for tungsten and steel the total systematic uncertainty is 0.3 ns at small radii and up to 0.8 ns at the largest radius due to the additional uncertainty originating from the geometrical differences of the two setups. The systematic uncertainties are summarized in Table \ref{tab:SystematicSummary}.

\begin{table}[t]
\caption{Summary of systematic uncertainties. Details are given in the text.}
\begin{center}
	\begin{tabular}{l|cc}
		source of uncertainty & uncertainty steel & uncertainty tungsten\\ \hline \hline
		relative timing between data sets &  \multicolumn{2}{c}{0.2 ns} \\	
		position uncertainty relative to beam axis & 0.2 -- 0.4 ns &0.2 -- 0.7 ns \\
		energy scale uncertainty & 0.1 -- 0.02 ns & 0.2 -- 0.02 ns\\
		time slewing correction & \multicolumn{2}{c}{0.1 ns} \\
		geometrical difference tungsten / steel &  \multicolumn{2}{c}{0 - 0.25 ns} \\
		normalization uncertainty of data sets &  \multicolumn{2}{c}{10\%} \\
		 \hline 
		 \multicolumn{3}{c}{combined uncertainties} \\ \hline \hline
		 data - MC vs energy & 0.2 ns & 0.3 -- 0.2 ns \\
		 data - MC vs radius & 0.3 -- 0.5 ns & 0.3 -- 0.7 ns \\
		 data steel - data tungsten vs radius & \multicolumn{2}{c}{0.3 -- 0.8 ns} \\	
		\hline		 
	\end{tabular}
\end{center}
\label{tab:SystematicSummary}
\end{table}

\subsection{Hadronic Shower Timing in Different Absorber Materials}

\begin{figure}[t]
        \centering
        \includegraphics[width=0.9\textwidth]{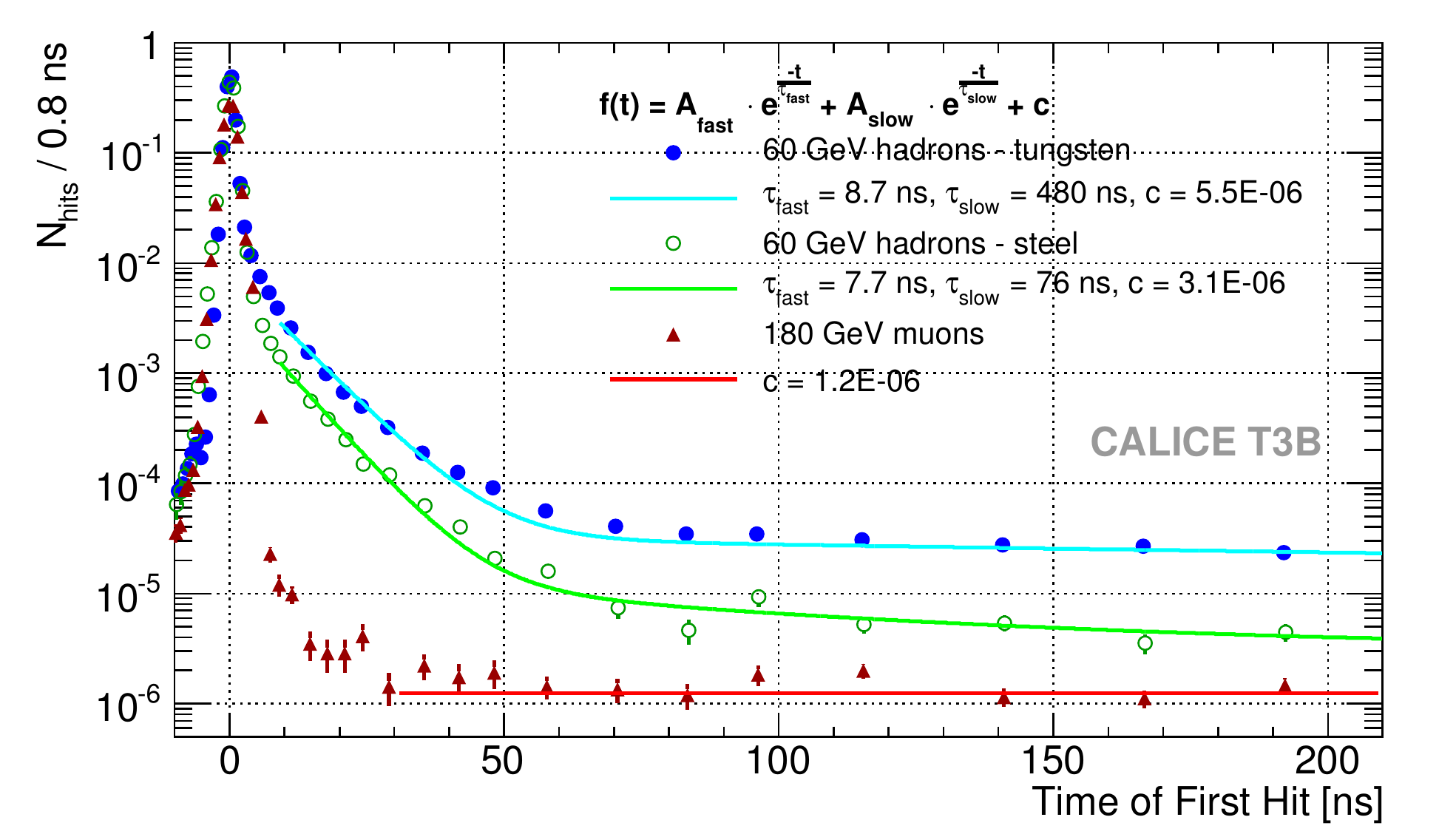} \\
        \includegraphics[width=0.9\textwidth]{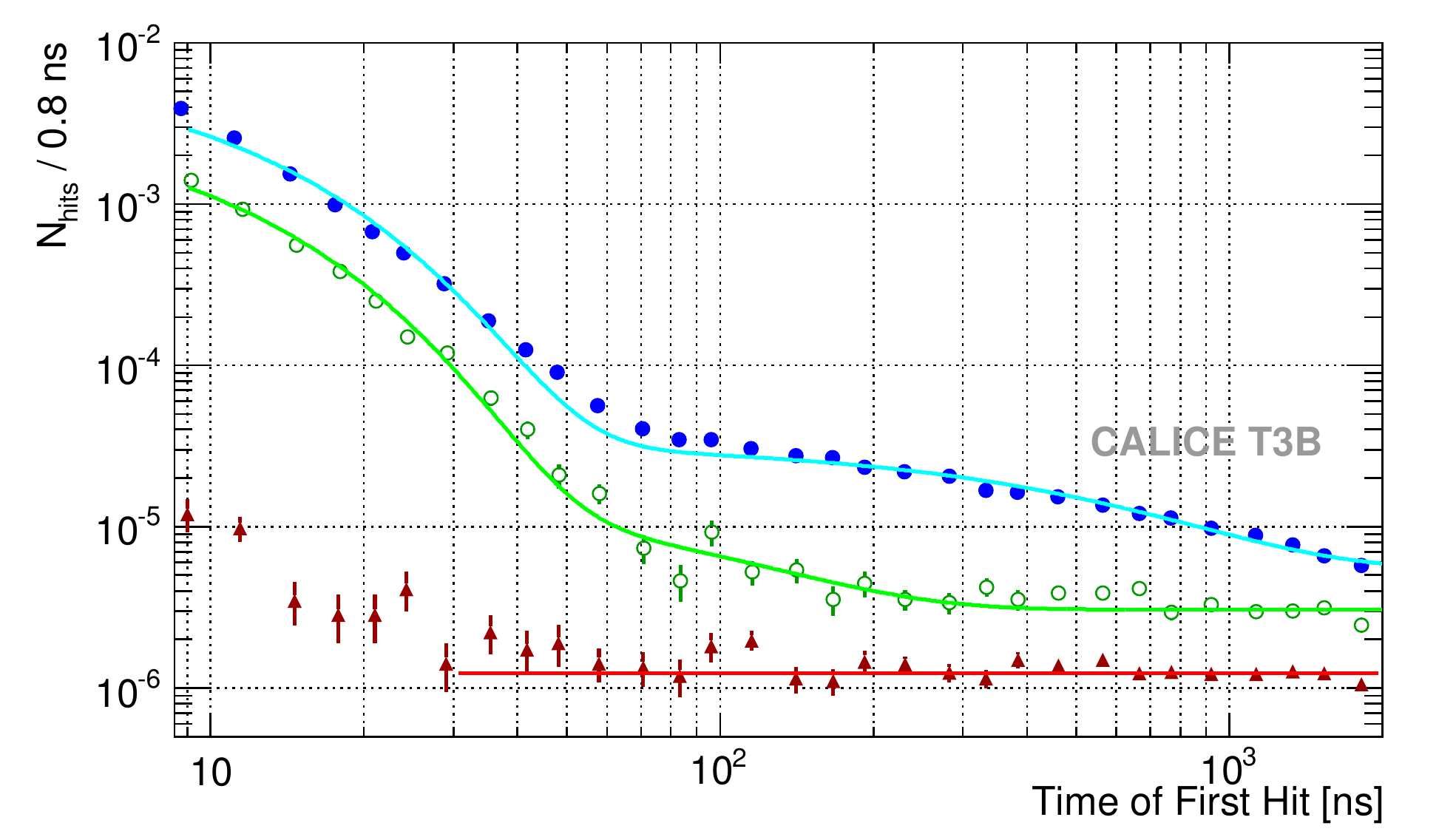} 
        \caption{Time of first hit distribution of muon data with steel absorbers and hadron data with steel and tungsten absorbers in a time range of $-10\,\text{ns}$ to $200\,\text{ns}$ (top). The histograms are normalized to the number of events in which at least one first hit could be identified and show the number of hits per T3B DAQ time bin of 0.8 ns. The same distributions are shown in a time range from $8\,\text{ns}$ up to $2000\,\text{ns}$ on a logarithmic time scale (bottom). Here, the peak of the distributions was excluded for better visibility. The lines show fits to the data, as described in the text.}
        \label{fig:T3BAnalysis:TofH:FitComponents}
\end{figure}

Within a hadronic cascade, absorber materials with high atomic number Z and higher neutron content are expected to release an increased number of evaporation neutrons. Such neutrons contribute substantially to delayed energy depositions predominantly by two mechanisms relevant at different times relative to the first interaction. The elastic scattering of evaporation neutrons with the active detector material is expected to contribute on time scales of a few ns to a few tens of ns, while neutron capture induces energy depositions with delays up to several $\mu$s due to the time-of-flight of low-energy neutrons and the lifetimes of unstable states. Thus, an increased late component of the showers in tungsten (Z = 74) is expected compared to steel (Z = 26), as discussed in more detail in Section \ref{sec:PhysicsIntro}. 

This expectation is confirmed by the data. Figure \ref{fig:T3BAnalysis:TofH:2DDistribution} (top) shows that all energy depositions initiated by impinging muons are concentrated in a small time window of a few ns around the trigger time (t=0). The isolated late hits in this figure are caused by SiPM noise and give a good assessment of the quality of the noise rejection criteria applied
for the analysis. Less than $0.05\,\%$ of all first hits are identified at a hit time later than
$8\,\text{ns}$. The situation is very different for the hadron data samples. In the case of steel, shown in Figure
\ref{fig:T3BAnalysis:TofH:2DDistribution} (bottom left), $1.2\,\%$ of the first hits are detected with a delay of more
than $8\,\text{ns}$. In the case of tungsten, presented in Figure \ref{fig:T3BAnalysis:TofH:2DDistribution} (bottom right),  the late shower component
is even more emphasized with $3.6\,\%$ of the first hits occurring delayed. While the largest part of a hadron shower
($99.92\,\%$ of the first hits) has decayed after $50\,\text{ns}$ in the case of steel data, the tungsten data exhibits a notable late
activity ($0.5\,\%$ of the first hits) even at times beyond $50\,\text{ns}$ after the particle impact.

In general, it is observed that the late shower activity is significantly larger for tungsten compared to steel data at all times. This is
shown in Figure \ref{fig:T3BAnalysis:TofH:FitComponents}, in which the time distribution of all first hits, normalized to the number
of events with T3B hits, is plotted. The number of hits is given in bins of 0.8 ns, the time binning provided by the T3B data accquision. The top part of the figure shows the distribution up to 200 ns after the trigger on a linear scale, while the bottom part shows the region from 8 ns to 2000 ns on a logarithmic scale. Similarly to the muon time distribution, the hadron data samples exhibit a quasi-instantaneous component
which contributes (due to the intrinsic time resolution of T3B, see \cite{Simon:2013zya} for details) only in a range of $-8$ ns to
8 ns. Additionally, in the case of hadron data, there is a fast shower component $\tau_{\text{fast}}$ contributing at intermediate times
which passes smoothly into a slow shower component $\tau_{\text{slow}}$ (at $t\approx50\ \text{ns}$). Therefore, for a first evaluation of the
involved physics processes, the late time development of hadron showers is fitted by a simple model which consists of the sum of two exponential
decays and a constant c, 
\begin{equation}
\frac{d\,N}{d\,t} \, \frac{1}{N_{\text{tot}}} = A_{\text{fast}} \cdot e^{(-\frac{t}{\tau_{\text{fast}}})} + A_{\text{slow}} \cdot e^{(-\frac{t}{\tau_{\text{slow}}})} + c, \label{eq:fit}
\end{equation}
where $N$ is the number of identified first hits, $N_{\text{tot}}$ is the total number of events with T3B hits, and $A_{\text{fast}}$ and $A_{\text{slow}}$ are the amplitudes of the fast and slow component, respectively.
The additional constant takes into account the random SiPM noise contribution and possible additional contributions with very long time constants,
which cannot be resolved by T3B due to the limited acquisition window. From laboratory measurements with the 15 SiPMs used in the T3B experiment, the noise contribution is expected to be of the order of 1 to 2 $\times\ 10^{-6}$ hits/ 0.8 ns, in agreement with the observed level of very late hits in the muon sample shown in Figure \ref{fig:T3BAnalysis:TofH:FitComponents}, fitted with a constant for times later than 30 ns. Variations in the run conditions (such as temperature or voltage variations) influence the occurrence of SiPM noise.

Since the fit is performed over six orders of magnitude on the vertical axis, a two-step approach is followed to ensure sensitivity to the low-amplitude late component. First, the slow component is fitted together with the constant in the range from $90\ \text{ns}$ to $2000\ \text{ns}$.  The parameters obtained are then fixed in the subsequent second fit with the full function as given in Equation \ref{eq:fit}, which is used to determine the fast component. With this procedure, a fast decay time of about $8\,\text{ns}$ is found for both hadron data samples ($7.7 \pm 0.1\ \text{ns}$ for steel, $8.7\pm 0.1\ \text{ns}$ for tungsten data), which is interpreted as signals originating from the scattering of MeV-scale evaporation neutrons in the active medium. The very late shower development differs considerably for steel and tungsten data. In the case of steel data, the slow decay occurs with a time constant of about $80\,\text{ns}$ ($76 \pm 1\ \text{ns}$). The contribution of this component to the total signal is less than 10\%  at times later than 290 ns,  where the constant $c$ dominates the distribution. For tungsten data, on the other hand, the slow decay time amounts to about $500\ \text{ns}$ ($480\pm 20\ \text{ns}$) and plays an important role (with a contribution of $>10\%$ to the total signal) up to the end of the investigated acquisition time at $2\ \mu\text{s}$. This is shown in Figure \ref{fig:T3BAnalysis:TofH:FitComponents} (bottom). From the figure it is also apparent that the fit function does not fully model the transition region from the fast to the slow component in tungsten in the region from $\sim$50 ns to $\sim$100 ns. This results in an increased $\chi^2/\text{NDF}$ of 18 for the full fit of the tungsten data, compared to 3.3 in the case of steel. Still, the fit gives a satisfactory description of the overall features of the observed time distribution for both absorbers. 

\begin{table}
\caption{Summary of the main results of the fit to the time of first hit distributions in Figure \ref{fig:T3BAnalysis:TofH:FitComponents}. The ratio of integrals is defined by Equation \ref{eq:RatioIntegrals}. See text for further details.}
\centering
\begin{tabular}{l|ccc}
fit parameter & steel & tungsten & ratio of integrals $R_i$\\
\hline
\hline
 $\tau_{\text{fast}}$ & $7.7 \pm 0.1\ \text{ns}$ & $8.7\pm 0.1\ \text{ns}$ & $2.3\pm0.5$ \\
 $\tau_{\text{slow}}$ & $76 \pm 1\ \text{ns}$ & $480\pm 20\ \text{ns}$ & $13.4\pm2.7$ \\
 \hline
 \hline
 & steel & tungsten & muons\\
 constant& $(3.06 \pm 0.08) \times 10^{-6}$ & $(5.48 \pm 0.19) \times 10^{-6}$ & $(1.24 \pm 0.03) \times 10^{-6}$  \\
 \hline
\end{tabular}
\label{tab:FitResults}
\end{table}

The highly emphasized late shower component of tungsten relative to steel data can be quantified by the ratio $R_i$ of the respective
amplitudes $A_{i}$ multiplied by the corresponding extracted time constants $\tau_{i}$, 
\begin{equation}
R_i = \frac{A_i^{\text{W}} \cdot \tau_i^{\text{W}}}{A_i^{\text{steel}} \cdot \tau_i^{\text{steel}}} \label{eq:RatioIntegrals}
\end{equation}
which is the ratio of the definite integrals of the exponential components from zero to infinity. A ratio $R_{\text{fast}} = 2.3$ is obtained for the fast component. For the slow component the ratio is found to be $R_{\text{slow}} = 13.4$. Table \ref{tab:FitResults} summarizes the main results of the fit.  
 
\begin{figure}[t]
        \centering
        \includegraphics[width=0.9\textwidth]{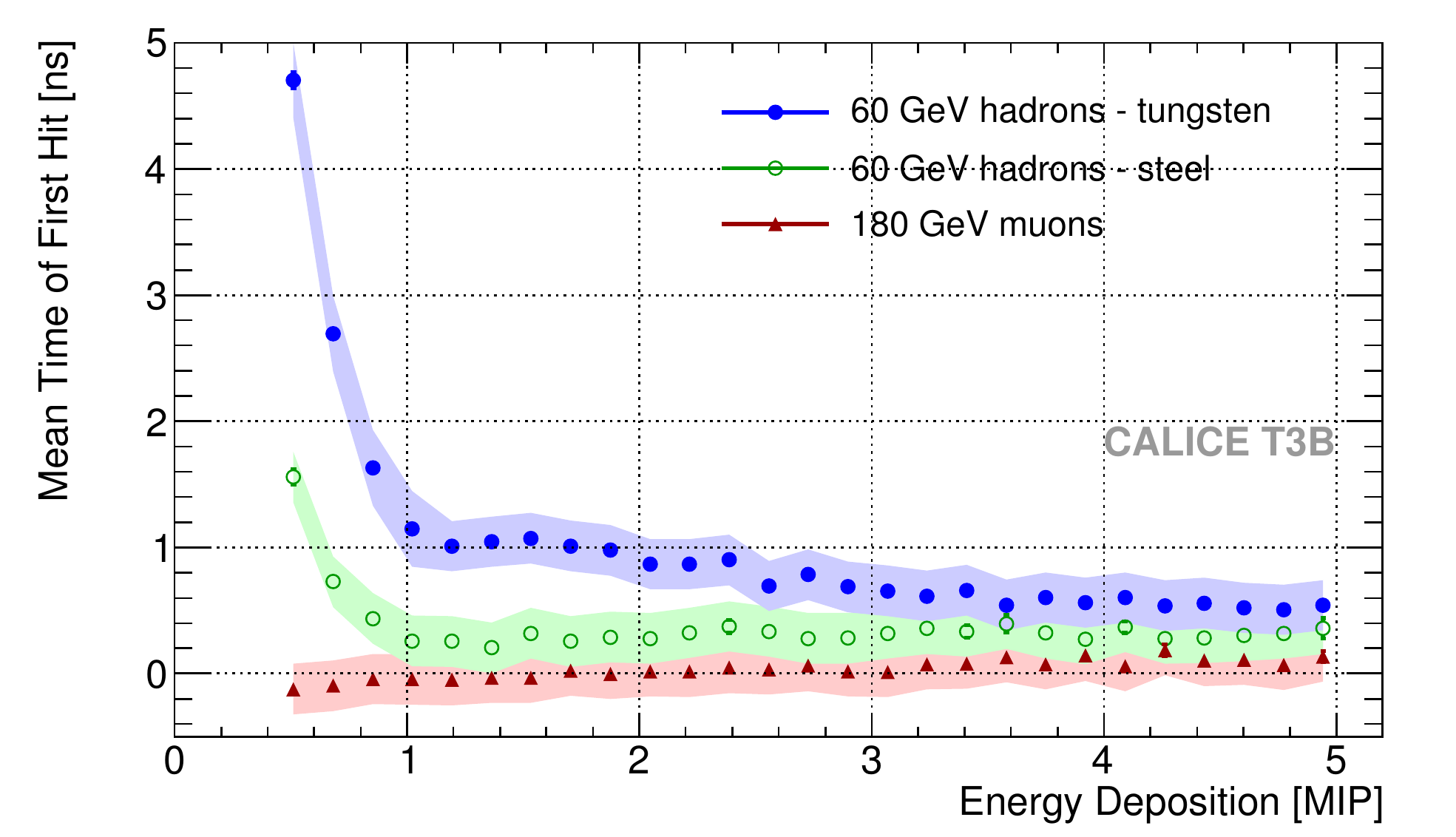}
        \caption{Energy dependence of the mean time of first hit for muon data with steel absorbers (red) and hadron data with steel (green) and tungsten (blue) absorbers. The bands show the systematic uncertainties.}
        \label{fig:T3BAnalysis:TofH:StWMu-Edistribution}
\end{figure}

The energy dependence of the timing of hits is studied in Figure \ref{fig:T3BAnalysis:TofH:StWMu-Edistribution}, which shows the mean time of
the first hit as a function of the hit energy for muon, steel and tungsten data. The mean TofH is always determined in the time range from
$-20\,\text{ns}$ and $200\,\text{ns}$ by taking the simple average of the time of all first hits in this time interval. Since the energy depositions induced by muons are prompt, no energy dependence is observed, as expected. The very slight slope of the time distribution visible in the figure is due to the modelling accuracy of the applied time slewing correction, and is smaller than the systematic uncertainties of the relative timing of the different data sets. 
For steel, the average hit time is slightly delayed by $1.6\ \text{ns}$ at $0.5\ \text{MIP}$, but this delay decreases quickly down to less
than $300\ \text{ps}$ for energies above $1\ \text{MIP}$. This demonstrates that higher energy deposits occur almost exclusively in the prompt part of the shower. The tungsten data, on the other hand, exhibits a significant delayed shower contribution at all energies. At $0.5\ \text{MIPs}$, the average delay is with $4.7\,\text{ns}$, about three times larger than for steel. At $\sim1\ \text{MIP}$, the mean TofH does still amount to more than $1.1\ \text{ns}$ and decreases down to $500\ \text{ps}$ at
$\sim5\ \text{MIP}$. Also here the importance of late energy deposits decreases with increasing energy, but contributions of delayed hits are found also at higher hit energies.

\subsection{Radial Dependence of Timing Profiles} 

\begin{figure}
\begin{center}
  \includegraphics[width=.9\textwidth]{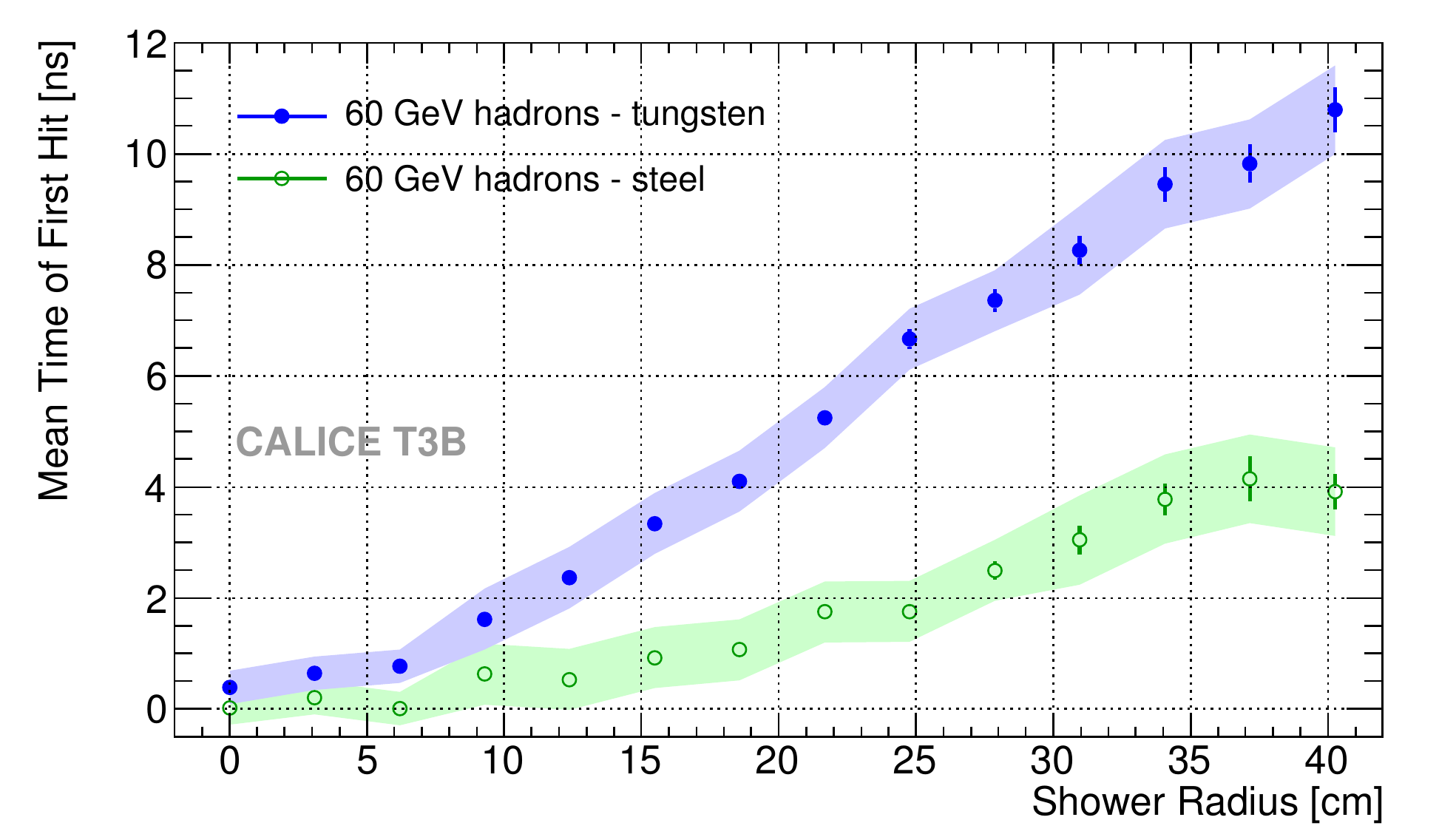}
	\caption{Radial shower timing profile of the mean time of first hit for hadron data with steel (green) and tungsten (blue) absorbers. The bands show the systematic uncertainties.}
	\label{fig:T3BAnalysis:MtofHvsRadius:WvsFe}
\end{center}
\end{figure}

The prompt shower contribution is dominated by electromagnetic subshowers and by relativistic particles, which are both concentrated along the shower axis. Low-energy neutrons on the other hand spread throughout the calorimeter, and thus contribute both close and far away from the beam axis. This is investigated by studying the lateral timing profile, given by the mean time of first hit as a function of radial distance from the shower axis, as shown in Figure \ref{fig:T3BAnalysis:MtofHvsRadius:WvsFe}. The  mean time of first hit exhibits an increase with increasing radius, consistent with the expectation of  a central core containing the majority of the shower energy, which is mainly deposited by prompt electromagnetic subshowers and relativistic hadrons, and a shower halo of mainly hadronic origin which, in addition to a prompt component, receives sizeable contributions from delayed signals, predominantly generated by neutron-induced processes. 

The increase of the mean time of first hit at larger radii is significantly more pronounced in tungsten than in steel. This is due to the larger yield of evaporation neutrons in tungsten, and due to the substantially shorter radiation length and the larger ratio of $\lambda_I$ to $X_0$ ($\sim27$ compared to $\sim10$)  for tungsten compared to steel.  At a radius of $\sim40\,\text{cm}$, the mean TofH is 2.8 times larger for tungsten than for steel ($10.8\ \text{ns}$ vs.\ $3.9\ \text{ns}$). In the shower center, this relative timing difference amounts to only $370\,\text{ps}$, which is of a similar order to the systematic uncertainties, as discussed in detail in Section \ref{sec:Systematics}.

\subsection{Comparison of Data to Monte Carlo Simulations}
\label{sec:MC}

\begin{figure}
\begin{center}
	\centering
	\includegraphics[width=0.495\textwidth]{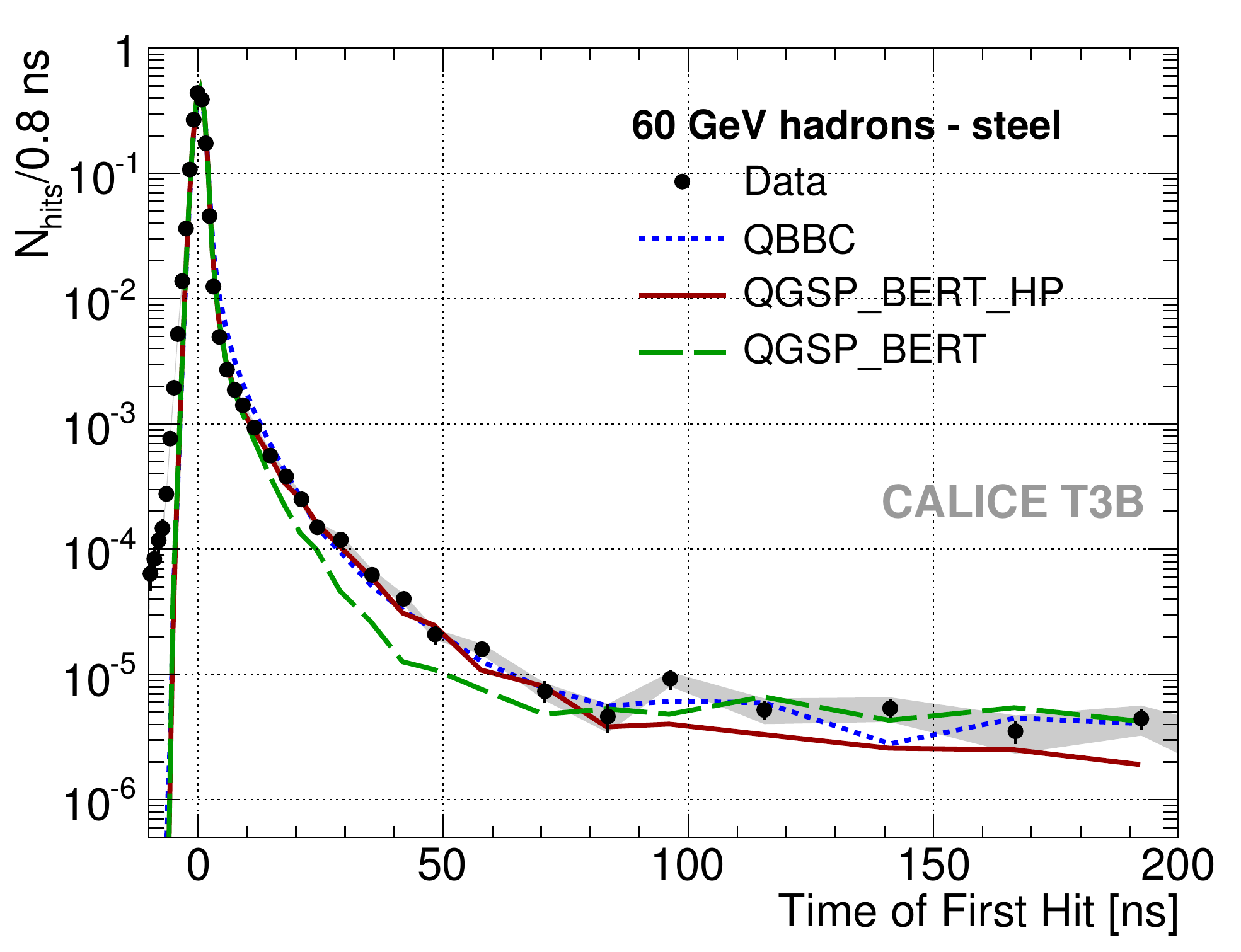}
	\includegraphics[width=0.495\textwidth]{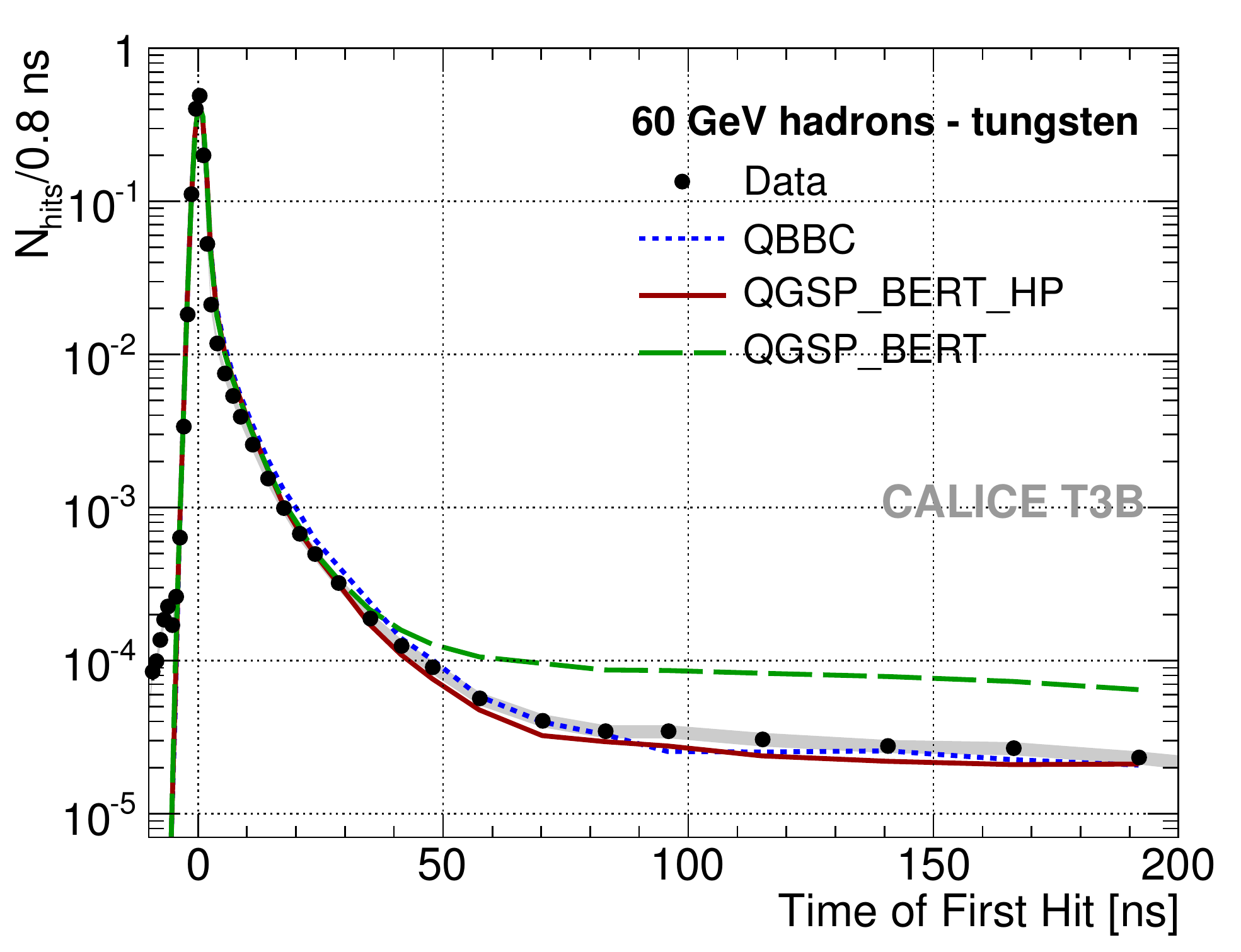} \\
	\caption{Comparison of the distribution of the identified hits in Monte Carlo and test beam data for hadrons in steel (left) and tungsten (right) for the time period from -10 ns to 200 ns. The grey band shows the systematic uncertainties.} 
	\label{fig:TimeDistributionMC} 
	\end{center}
\end{figure}
\begin{figure}
\begin{center}
	\includegraphics[width=0.495\textwidth]{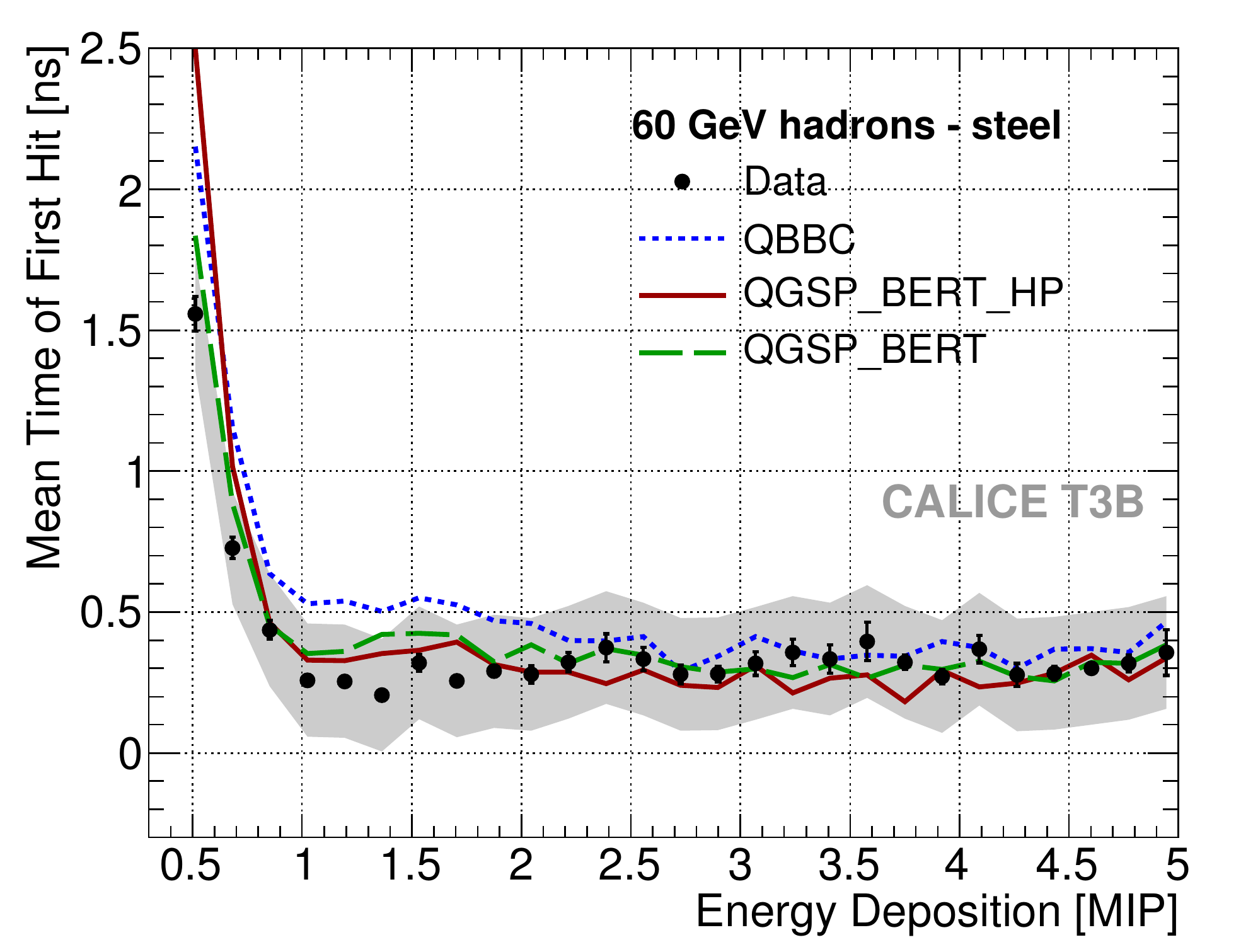}
	\includegraphics[width=0.495\textwidth]{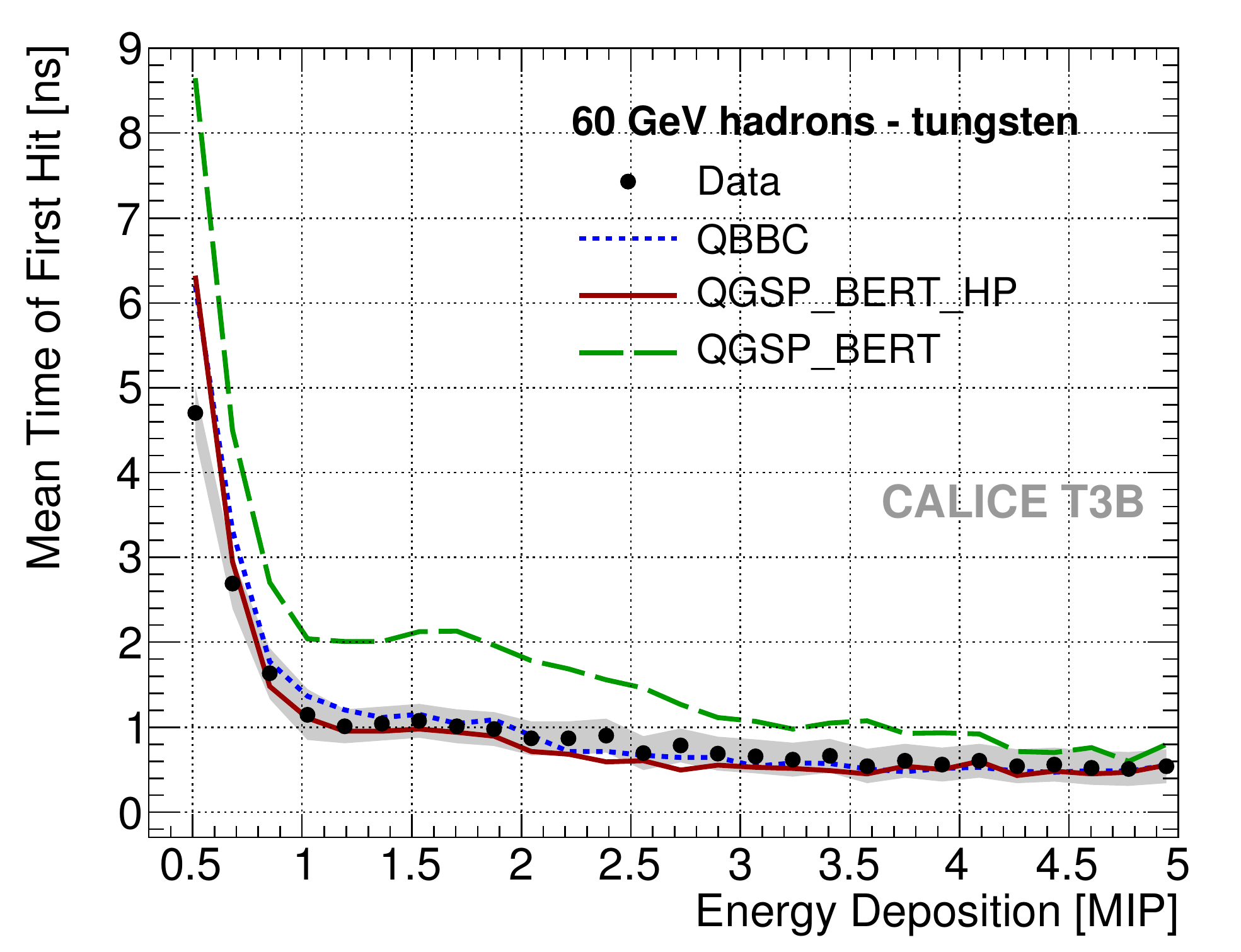} \\
	\caption{Comparison of the energy dependence of the mean time of first hit of Monte Carlo and test beam data for hadrons in steel (left) and tungsten (right) absorbers. The grey band shows the systematic uncertainties.} 
	\label{fig:T3BAnalysis:MtofHvsRadius:DataVsSim-EDep}
	\vspace{0.25cm}
	\includegraphics[width=0.495\textwidth]{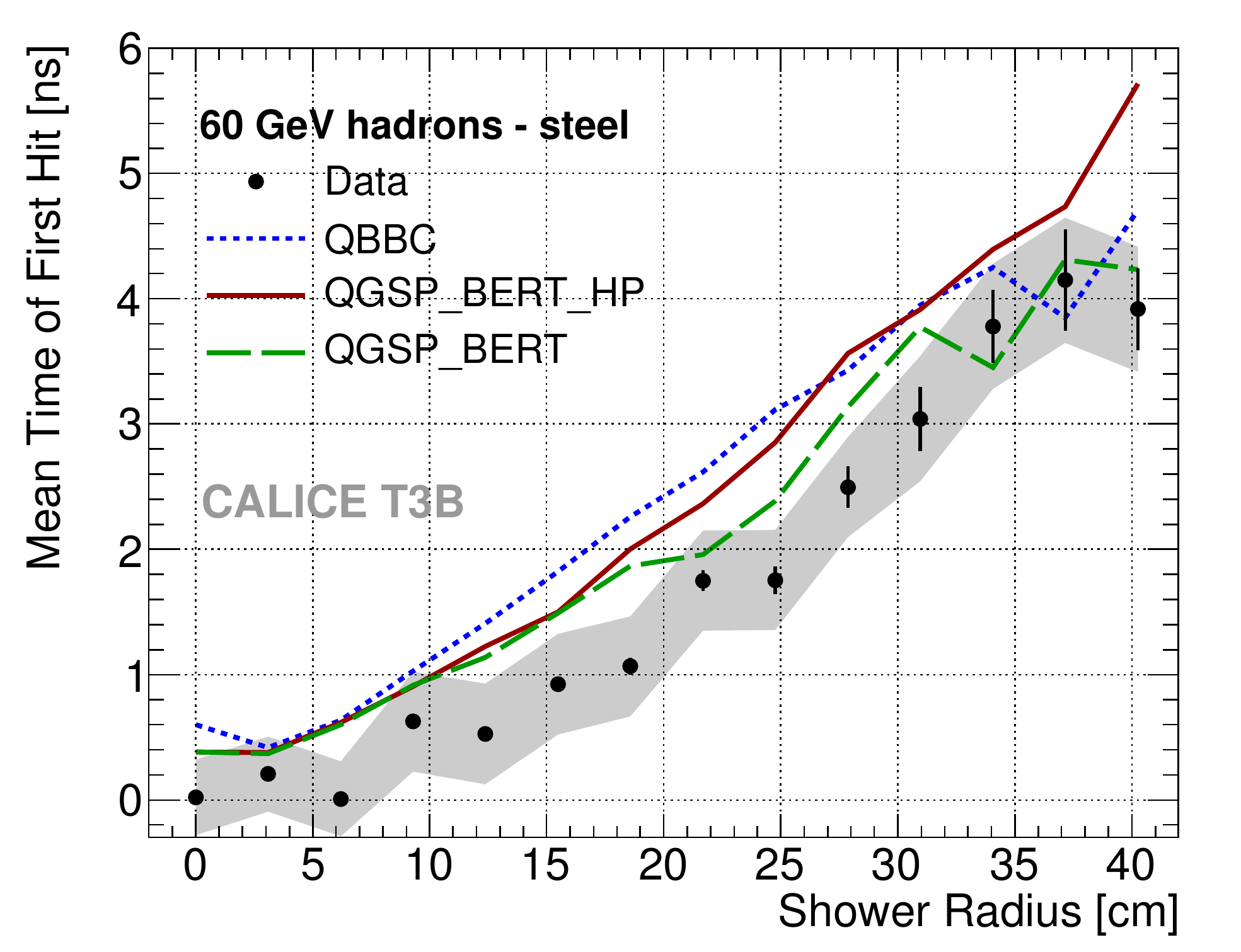}
  	\includegraphics[width=0.495\textwidth]{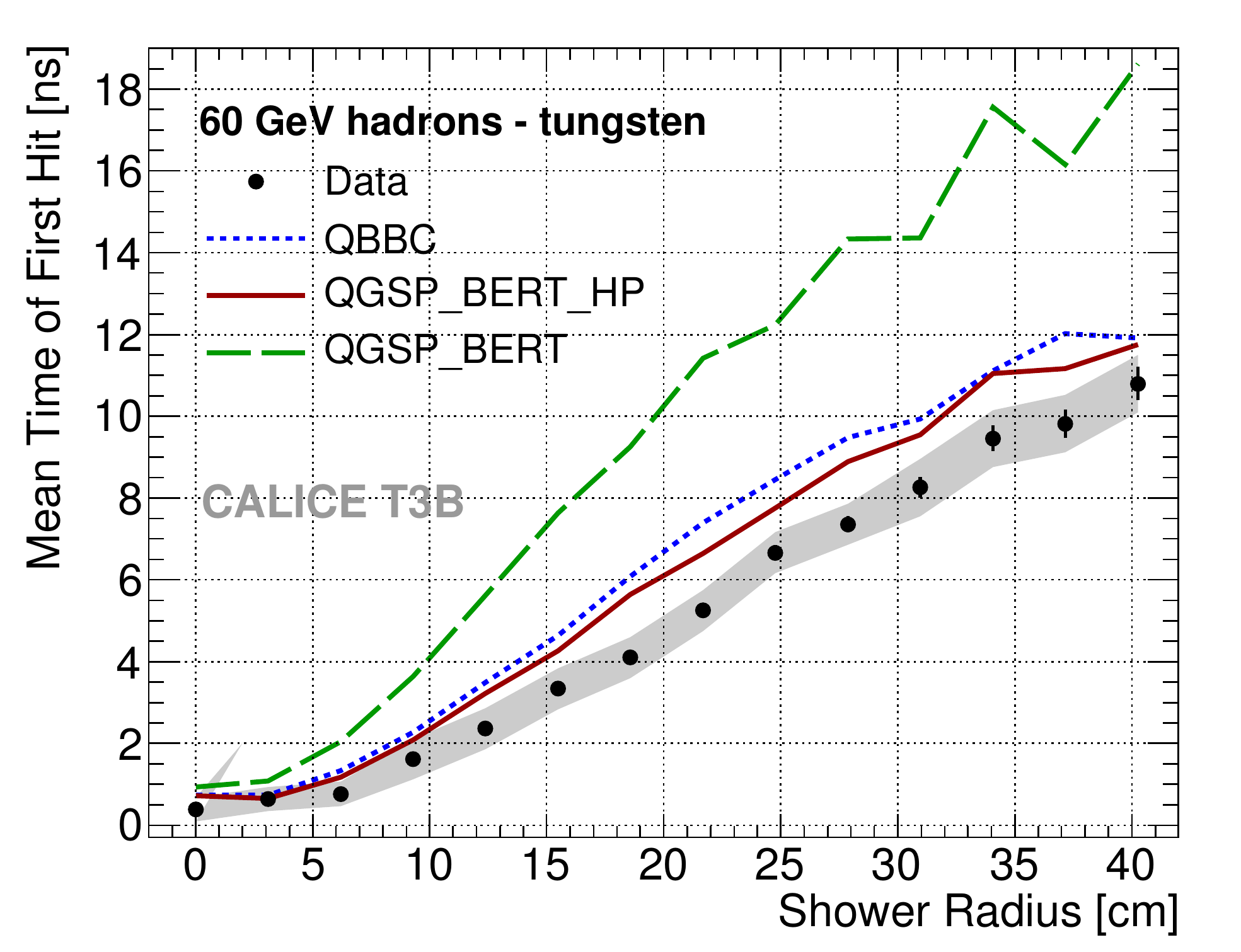} 
  	\caption{Comparison of the radial profile of the mean time of first hit of Monte Carlo and test beam data for hadrons in steel (left) and tungsten (right) absorbers. The grey band shows the systematic uncertainties.}
  	\label{fig:T3BAnalysis:MtofHvsRadius:DataVsSim-RadialDep}
\end{center}
\end{figure}

To determine the accuracy of shower timing in  \geant, the T3B data are compared to simulations based on different hadronic physics models, as introduced in Section \ref{sec:Sim}.  Figure \ref{fig:TimeDistributionMC} shows the time distribution of the first hits in steel and tungsten absorbers compared to the three physics lists. While QBBC and QGSP$\_$BERT$\_$HP reproduce the distribution well for both absorbers, QGSP$\_$BERT shows some discrepancy with the data. In steel, the shower activity in the intermediate time period from 10 ns to 60 ns is slightly underestimated, while in tungsten the late component $> 50$ ns is overestimated by up to a factor of four. 

Figure \ref{fig:T3BAnalysis:MtofHvsRadius:DataVsSim-EDep} shows the energy dependence of the mean time of first hit, which is well reproduced by all hadronic models considered here in steel, but deviates substantially for QGSP$\_$BERT in tungsten.  Although the functional form is similar to the one from data, the mean TofH turns out to be between $2\,\text{ns}$ and $0.5\,\text{ns}$ too large over a wide hit energy range of $0.4\,\text{MIP}$ to $3.5\,\text{MIP}$, showing that the physics list without high precision neutron tracking produces too many late energy depositions in particular in the lower energy range below 3 MIP (2.5 MeV). 

The radial timing profile of the shower, shown compared to simulations in Figure \ref{fig:T3BAnalysis:MtofHvsRadius:DataVsSim-RadialDep}, further confirms these observations. For steel, all physics lists agree with each other and with data within 1 ns. For tungsten, QGSP$\_$BERT overestimates the delayed shower contribution, and with that the mean time of first hit at all radii. The discrepancy is seen to increase with increasing distance from the shower axis. While the difference to data in the mean TofH amounts to only $2.0\ \text{ns}$ at a radius of $9.2\ \text{cm}$, it increases up to $7.8\ \text{ns}$ in the outer shower region at $40.2\ \text{cm}$. For the high precision lists, in general the  timing profile agrees well with data, with a slight overestimation of the TofH at the 1 -- 2 ns level for radii larger than 10 cm.


\section{Conclusion}
\label{sec:Conclusion}

The time structure of hadronic showers, and the level of accuracy with which it can be simulated in \geant, is highly relevant for calorimeters at future collider experiments. This applies in particular in conditions with high background levels and high repetition rates, such as at CLIC, where tungsten is considered for the absorber material of the hadron calorimeter. The T3B experiment studies the time structure of hadronic showers on a statistical basis with large event samples in hadron calorimeters with steel and tungsten absorbers collected in conjunction with the CALICE imaging calorimeter prototypes. T3B is based on 15 small scintillator tiles with SiPMs and fast USB oscilloscope readout arranged in a strip to provide full radial sampling of the showers with high granularity and sub-ns time resolution over an acquisition window of 2.4 $\mu$s.  

The data show late components of hadronic showers, which are substantially more pronounced in tungsten than in steel. The late component is predominantly concentrated at lower hit energies, but also extends up to several MIP-equivalents in tungsten. The importance of the late energy depositions increases with increasing distance from the shower axis, due to the wide lateral spread of late activity driven by neutrons in contrast to the more concentrated evolution of the prompt electromagnetic subshowers and relativistic hadrons. The comparison of detailed detector simulations with the data shows that the time structure is generally quite well modelled in steel. In contrast, the tungsten data is only reproduced by models with a dedicated treatment of low-energy neutrons, such as the high-precision neutron package in QGSP\_BERT\_HP, or alternative implementations used in the QBBC model. The QGSP\_BERT physics list, which is widely used for LHC and linear collider detector simulations, substantially overestimates the amount of late energy depositions in tungsten.

\section*{Acknowledgements}
\label{sec:Acknowledgements}

We gratefully acknowledge help by the technical staff at several CALICE institutes for their help with the WAHCAL / T3B, and SDHCAL / T3B test beams. We also gratefully acknowledge the DESY and CERN managements for their support and hospitality, and their accelerator staff for the reliable and efficient beam operation. The authors would like to thank the RIMST (Zelenograd) group for their help and sensors manufacturing. This work was supported by the European Commission under the FP7 Research Infrastructures project AIDA, grant agreement no. 262025; by the Bundesministerium f\"ur Bildung und Forschung, Germany; by the the DFG cluster of excellence `Origin and Structure of the Universe' of Germany; by the Helmholtz-Nachwuchsgruppen grant VH-NG-206; by the BMBF, grant no. 05HS6VHS1; by the Russian Ministry of Education and Science contracts 4465.2014.2 and 14.A12.31.000 and the Russian Foundation for Basic Research grant 14-02-00873A; by MICINN and CPAN, Spain; by CRI(MST) of MOST/KOSEF in Korea; by the US Department of Energy and the US National Science Foundation; by the Ministry of Education, Youth and Sports of the Czech Republic under the projects AV0 Z3407391, AV0 Z10100502, LC527 and LA09042 and by the Grant Agency of the Czech Republic under the project 202/05/0653; by the National Sciences and Engineering Research Council of Canada; and by the Science and Technology Facilities Council, U.K.

\bibliography{bibliography}

\end{document}